\documentclass[manuscript, authorversion]{acmart}

\AtBeginDocument{%
  }

\setcopyright{acmlicensed}
\copyrightyear{2018}
\acmYear{2018}
\acmDOI{XXXXXXX.XXXXXXX}
\acmConference[Conference acronym 'XX]{Make sure to enter the correct
  conference title from your rights confirmation email}{June 03--05,
  2018}{Woodstock, NY}
\acmISBN{978-1-4503-XXXX-X/2018/06}

\usepackage{multirow}
\usepackage{tabularx}
\usepackage{subcaption}
\usepackage{graphicx}
\usepackage[capitalize, noabbrev]{cleveref}
\usepackage[export]{adjustbox}

\usepackage{tabularray}
\UseTblrLibrary{booktabs}

\begin{document}

\title[Functional Flexibility in Generative AI Interfaces]{Functional Flexibility in Generative AI Interfaces: Text Editing with LLMs through Conversations, Toolbars, and Prompts}

\author{Florian Lehmann}
\orcid{0000-0003-0201-867X}
\email{florian.lehmann@uni-bayreuth.de}
\affiliation{%
  \institution{University of Bayreuth}
  \streetaddress{Universit\"atsstra{\ss}e 30}
  \city{Bayreuth}
  \country{Germany}
  \postcode{95447}
}

\author{Daniel Buschek}
\orcid{0000-0002-0013-715X}
\email{daniel.buschek@uni-bayreuth.de}
\affiliation{%
  \institution{University of Bayreuth}
  \streetaddress{Universit\"atsstra{\ss}e 30}
  \city{Bayreuth}
  \country{Germany}
  \postcode{95447}
}

\renewcommand{\shortauthors}{Lehmann and Buschek}

\newcommand{\lastaccessed}{\textit{last accessed 10.10.2024}}

\definecolor{FloriansColor}{rgb}{0,0.3,0.9}
\newcommand{\florian}[1]{\textsf{\textbf{\textcolor{FloriansColor}{[Florian: #1]}}}}

\definecolor{DanielsColor}{rgb}{0.9,0.6,0.1}
\newcommand{\daniel}[1]{\textsf{\textbf{\textcolor{DanielsColor}{[Daniel: #1]}}}}

\definecolor{colorrevision}{HTML}{000000} %
\definecolor{colorrevisiontwo}{HTML}{000000}%
\definecolor{colorrevisiontwoflo}{HTML}{000000}
\definecolor{colorrevisionthree}{HTML}{000000}%
\definecolor{colorrevisionthreeflo}{HTML}{000000}%

\newcommand{\revision}[1]{\textcolor{colorrevision}{#1}}
\newcommand{\revtwo}[1]{\textcolor{colorrevisiontwo}{#1}}
\newcommand{\revtwoflo}[1]{\textcolor{colorrevisiontwoflo}{#1}}
\newcommand{\revthree}[1]{\textcolor{colorrevisionthree}{#1}}
\newcommand{\revthreeflo}[1]{\textcolor{colorrevisionthreeflo}{#1}}

\newenvironment{longrevision}{%
  \par 
  \begingroup
  \color{colorrevision}%
}{%
  \par
  \endgroup
}

\frenchspacing

\begin{abstract}

Many recent user interfaces (UIs) promise great flexibility via access to Large Language Models (LLMs) through prompting. This shifts efforts in accessing functionality from developers to users. However, is a maximum of functional flexibility always beneficial? We find: no -- \revthreeflo{and develop \revtwoflo{Functional Flexibility} as a concept by reflecting on} our long-term exploration of integrating LLMs into a text editor\revtwo{:} In a prompt-elicitation survey, people (N=121) envisioned elaborate prompts for delegating flexible tasks to AI. In contrast, when interacting with a prototype (N=10), users preferred short command-like prompts, as prompting repeated functionality is tedious. Another prototype offered such commands in a toolbar, in addition to flexible prompting, leading users (N=12) to dynamically work with both. We \revtwoflo{contribute Functional Flexibility} as a theoretical construct with analytical, critical, and constructive value. Overall, we encourage designing UIs for generative AI that go beyond free text prompting by explicitly considering functional flexibility in design.

\end{abstract}




\maketitle

\begin{figure*}[t]
    \centering
    \includegraphics[width=\textwidth]{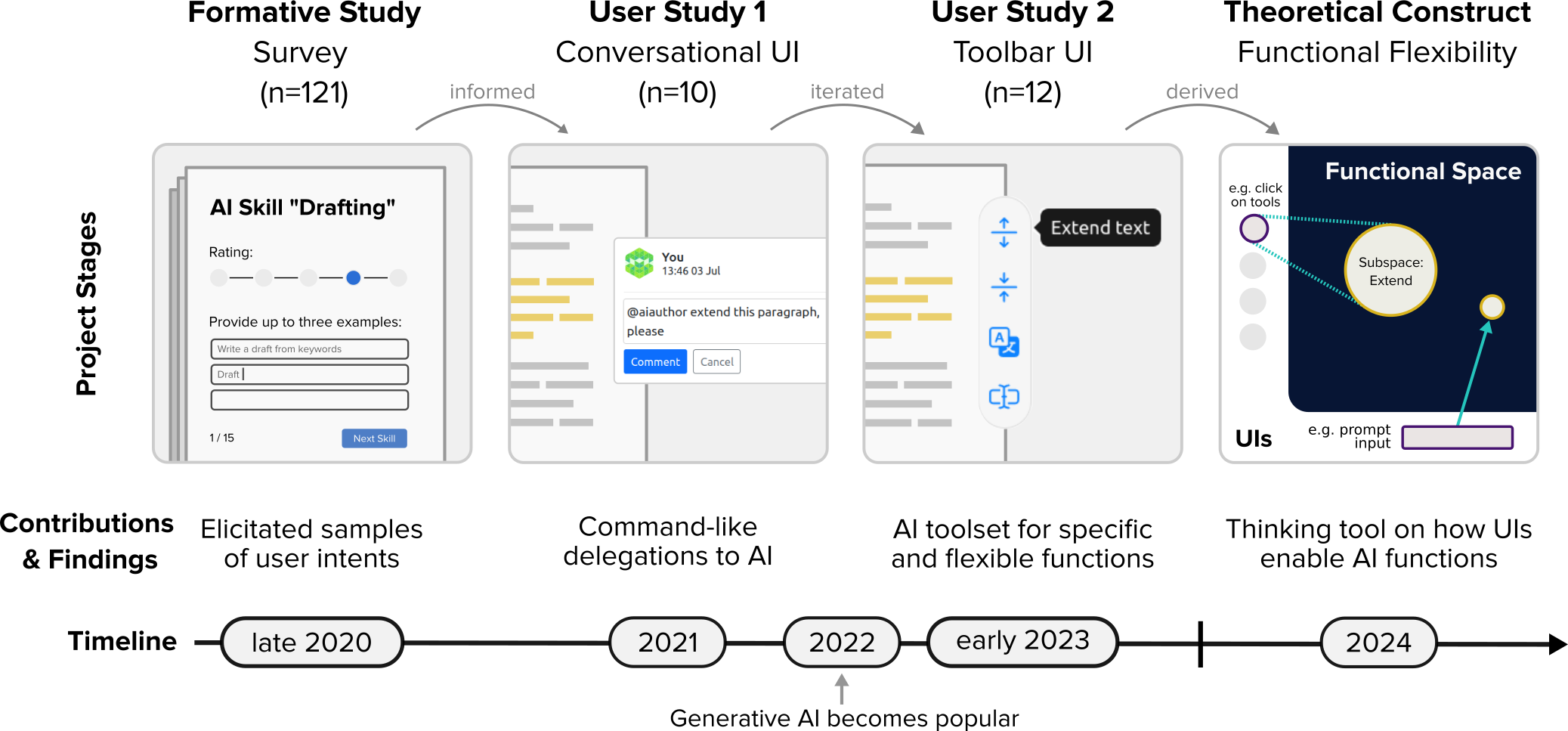}
    \caption{We investigated text editing with generative text models over four years from 2020 to 2024, covering the moment when generative AI became widely popular. This research effort has four parts (left to right): (1) a formative survey to explore people's text editor usage and to elicit how users would delegate tasks to AI (N=121), (2) a user study with a prototype integrating conversational AI (N=10), (3) an alternative prototype offering AI tools, again tested in a user study (N=12), (4) a theoretical view on UIs and AI with a focus on functional flexibility. In summary, we contribute a set of elicited samples of user intents, findings from our studies about AI use in text editors, and a theoretical perspective on how UIs shape functional flexibility.}
    \label{fig:framing}
    \Description{This figure shows the timeline of our project that covered four years from 2020 to 2024, including the moment when generative AI became widely popular. The timeline visualizes the four parts of our investigation from left to right: 1) formative study, 2) user study 1 on conversational UIs, 3) user study 2 on toolbar UIs, and 4) the theoretical construct synthesized from our findings and related work.}
\end{figure*}

\Description{This table gives an overview of the participants, including columns that describe the writing project, regular audiences, writing proficiency, English proficiency, age, gender, and occupation for each participant.}

\section{Introduction}

\revision{Prompting is in focus in current HCI research (e.g. \cite{lee2024designspace, subramonyam_bridging_2024, arawjo_chainforge_2024, suh_luminate_2024}) but also in industry. Microsoft, for example, has integrated chatbots based on large language models (LLMs) into various tools to extend their functionality (e.g. see Figure 4 in \cite{tankelevitch_metacognitive_2024}). At the same time, research showed that prompting is demanding \cite{zamfirescu-pereira_why_2023, tankelevitch_metacognitive_2024}, and \citet{morris_prompting_2024} recently argued that such prompting interfaces might not be optimal for end users.}

\revision{We concretely challenge the current focus on prompting with the terminology of \textit{functional flexibility} that we introduce in detail in this paper: With prompts, users express their intents that serve underlying goals. For example, a prompt for gaining inspiration might say ``Give me inspiration for an essay about recent advances in technology''. Conceptually, the user's intended function of this prompt is to ``inspire'' the user. LLMs, such as generative pre-trained transformers (GPTs) \cite{brown_language_2020, radford_language_2019} are \textit{flexible} in executing such intent-driven functions, since they posses an almost unlimited number of capabilities for generating or transforming text. That is, they offer a broad \textit{functional space}. Thus, the distinctive value of these models for users is their \textit{functional flexibility}. }

\revision{In this view, the role of user interfaces (UIs) is to make this functional flexibility available to end users. Currently, entering prompts as text is the predominant UI choice for this (e.g. ChatGPT). This grants users a \textit{maximum} of functional flexibility, because free text input enables them to specify \textit{any} request and access a broad range of model outputs (with some degree of quality).} However, maximizing functional flexibility might not always be in the best interest of the users, for example, when it comes at the expense of knowing how to achieve something (cf.~\cite{tankelevitch_metacognitive_2024}) \revision{-- e.g. an empty chat box gives little guidance on what the system can do.}

We thus argue that it is important \revision{for HCI researchers and practitioners} to understand which level of functional flexibility is given to users through various UI designs, to be able to make this an \textit{explicit} design choice. Toolbar UIs, for example, could offer specific AI functions to users as buttons, while text input UIs offer access to more flexible AI functions. \revision{A UI might also provide both.} In this way, applications could \revision{(also)} offer a set of pre-determined AI functions to the user without \revision{demanding} a user prompt. \revision{This example of} AI tools shows that there are more possibilities for interaction with AI than conversing with and prompting models. Thus, when designing UIs for AI, we see the need to better understand how AI can be integrated into users' actual workflows, considering functional flexibility.

\revtwoflo{The core contribution of this paper is our concept on Functional Flexibility. We propose and develop this concept in the context of interaction with LLMs.} We arrive at this through reflection on our multi-faceted research efforts on text editing with LLMs, spanning 2020 to 2024. %
Concretely, \revthreeflo{our perspective is shaped by interpreting} insights from probing user expectations and prompting behaviour both before and after the widespread use of prompting LLMs (i.e. release of ChatGPT) (see \cref{fig:framing}). 

This long-term perspective allowed us to see contrasts and variations in user behaviour and model capabilities, \revthreeflo{pointing towards} the role of UIs as shaping functional flexibility and users' exploration of the functional spaces provided by LLMs. %
In more detail, \revthreeflo{we report on three empirical studies that shaped our theory-building:} (1) our formative survey (N=121) elicited how users envision delegating writing tasks to AI. This informed our design of a conversational UI (CUI) prototype for text editing with AI. (2) A mixed-methods technology probe study (N=10) revealed that people often entered very short instructions. (3) Our second prototype then provided these directly as shortcuts in a toolbar UI, in addition to prompting, evaluated in a second study (N=12).

\revthreeflo{These studies revealed findings that we interpret to develop the concept of Functional Flexibility:} We found that -- unbiased by exposure to current LLM-based chatbots -- users optimized for command-like instructions in actual interactions, %
although people had envisioned longer prompts in the survey. This changed when adding a toolbar for common functions: The resulting mix of toolbar and free text prompting was used to dynamically switch between pre-specified AI functions and investing in longer prompts for flexible functions. %
Moreover, with the toolbar, users integrated AI functions closely into text editing for specific tasks and chained functions to reach their goals.
\revision{With \textit{Functional Flexibility}, we brought together} these insights in a new theoretical construct as a thinking tool for researchers and designers. Inspired by Generative Theories of Interaction \cite{beaudouin-lafon_generative_2021}, \revision{we compiled a set of} analytical, critical, and constructive \revision{questions to reflect on our theoretical construct. Finally, we deconstruct and discuss a recent example of a feature-rich human-AI interface from related work \cite{siddiqui_scriptshift_2025} to demonstrate the applicability of our theory.} \revthreeflo{Beyond this, we intend the concept to guide -- and to be validated and refined by -- further empirical work, such as by applying its considerations in the design of future interfaces.}

Overall, our multi-faceted exploration of human-AI interaction encourages researchers, designers, and developers of interactive AI systems to move beyond today's predominantly conversational UIs for generative AI. Concretely, we \revision{recommend doing} so by analyzing and reflecting explicitly on how different UIs shape users' access to the functional space of generative AI models. 

The paper is structured as follows: First, we motivate functional flexibility as a fundamental concept in light of the literature (\cref{sec:related_work}). Second, we report on our empirical work: a prompt-elicitation survey (\cref{sec:survey}), and two subsequent user studies with prototypes as technology probes, which focused on a conversational UI (\cref{sec:study1}) and its evolution (\cref{sec:prototype_design_changes}), leading to a version with toolbar plus prompting (\cref{sec:study2}). We then synthesize our \revthreeflo{reflections on these studies} with contributions from related work and describe in detail our theoretical construct of functional space and functional flexibility (\cref{sec:functional_space}). Throughout this and the following discussion (\cref{sec:discussion}), we consider functional flexibility from analytical, critical, and constructive perspectives \cite{beaudouin-lafon_generative_2021}, along with additional findings.

We release our prototype and material in this project repository to support future research: \url{https://osf.io/qh46n}

\section{Related Work}\label{sec:related_work}

Many recent intelligent writing tools use LLMs based on transformer models \cite{vaswani_attention_2017}. Task-specific models serve one purpose such as translating \cite{tiedemann_opus-mt_2020} and summarizing text \cite{raffel_exploring_2023}. General purpose models \cite{devlin_bert_2019, radford_language_2019, liu_roberta_2019, brown_language_2020}, such as GPTs, generalize well across language tasks, leveraging huge training corpora. Before such models were available, research on writing support systems introduced tools surrounding the writing process, for example, to enhance consistency \cite{han_textlets_2020}, facilitate planning \cite{lu_inkplanner_2019}, and support creativity \cite{maiden_making_2018}. 
With such models, many intelligent writing tools now generate suggestions and involve prompting~\cite{lee2024designspace}. For implementing our prototypes, we combined task-specific models for summarizing and translating text, as well as a general-purpose models for generating text.

\subsection{Writing with generated suggestions}

Several studies examined the impact of writing with AI-generated text suggestions:
For example, integrating sentence-level suggestions in emailing is more demanding than message-level suggestions yet users perceive a higher sense of agency \cite{fu_comparing_2023}. Presenting multiple suggestions instead of only one is valued for inspiration \cite{buschek_impact_2021}. Writing with suggestions on mobile devices resulted in longer text but higher perceived authorship than editing AI-generated text \cite{lehmann_suggestion_2022}. Writing with text suggestions can influence users' views \cite{jakesch2023chi} and lead to more predictable text \cite{arnold_predictive_2020}. Suggestions can be shown constantly \cite{buschek_impact_2021}, after inactivity \cite{bhat_interacting_2023} or on user request \cite{dhillon_shaping_2024, lee_coauthor_2022}. Overall, studies such as these show that writing with phrase suggestions influences how users write and perceive text. \citet{bhat_interacting_2023} extended the cognitive process model of writing accordingly. %

Our prototypes let participants delegate writing tasks to AI, and they could then decide whether to include the results into their draft, in three ways (replace, append, copy) -- and further edit it in all cases.

\subsection{Writing with prompts}

Prompting is a common interaction technique for delegating tasks to LLMs. Related work has found positive effects, for example, that prompting during writing is used to gain inspiration and explore topics \cite{dang_choice_2023}, and to create own creative writing tools \cite{swanson_story_2021}. %
However, prompting comes with challenges \cite{zamfirescu-pereira_why_2023}, including debugging and evaluating effectiveness \cite{jiang_prompt-based_2022}, and added meta-cognitive demands \cite{tankelevitch_metacognitive_2024}. %
Despite these challenges, when prompting is integrated into text editors, it is typically considered a tool for creatives~\cite{lee2024designspace}: For example, \citet{yuan_wordcraft_2022} built an editor with generative capabilities accessed via prompting in a sidebar and conducted a study with novelists. Their tool increased writers' engagement, and users found it more helpful compared to only using AI text continuations. 

Similar to their work, we created an editor with generative capabilities. %
We restricted AI to a low level of initiative and let participants use different UIs (conversational UI, toolbar UI, prompting) to control text generation. Next, we give an overview of the fundamental concepts of initiative and control for designing human-AI interaction, and suggest that functional flexibility is one such concept.

\subsection{Fundamental concepts in designing for human-AI interaction}

Two fundamental concepts in designing for human-AI interaction are initiative and control. In mixed-initiative applications \cite{horvitz_principles_1999}, AI can take actions on its own. For example, related work investigated agent-initiated interactions with voice assistants \cite{reicherts_may_2021}, initiative as a design parameter \cite{clark_creative_2018}, and evaluated a collaborative moodboard tool with different levels of initiative and control \cite{koch_imagesense_2020}. While rarely explicitly defined in these papers, initiative is typically understood as starting and carrying out actions, while control concerns adjusting parameters and constraining system functionalities. 

Control is an important aspect in interaction with generative AI, such as for text generation \cite{lee2024designspace}. The sense of control is an often discussed aspect in HCI research \cite{bennett_how_2023}. %
Technically, such control can be facilitated by fine-tuning models for specific tasks \cite{zhong_fiction-writing_2023} or with prompting techniques, such as chaining prompts \cite{wu_ai_2021}. Related, \citet{singh_where_2022} reflect on the users' desire to steer the semantics of what is generated by LLMs. This underlines the importance of designing human-AI interaction in a way that addresses user desires, as users want to reach a goal by integrating LLM-based features into their workflows (i.e. reaching a user-expected function). 

We suggest that the flexibility of such functionality is another fundamental concept in designing for human-AI interaction. Concretely, to design effective interfaces, we need to understand which functions users expect from generative applications and focus on designing UIs for accessing these functions in the LLM's space of possible functions. In this paper, we investigate this as a fundamental concept across multiple studies, with an emphasis on the role of the UI. %

\subsection{Summary}

In summary, the functional flexibility of interactive generative models is currently overlooked as a distinct concept, as is evident, for example, from the absence of such a concept in comprehensive and widely referenced design guidelines for interaction with AI~\cite{weisz_design_2024, amershi_guidelines_2019}. Today, UI and interaction design for LLM-based systems often provide a maximum of the model's functional space to the user through free prompting. This maximizes functional flexibility but also shifts the task of identifying and determining functions from designers and developers to the end-user. This can be challenging for users, since giving effective instructions to an AI model is often difficult and cognitively demanding \cite{tankelevitch_metacognitive_2024, zamfirescu-pereira_why_2023}. In conclusion, we diagnose that, amidst the promises of prompting, there currently is no explicit consideration of the provided functional flexibility when designing interactive LLM-based systems. This motivates our investigation and perspective in this paper.

\section{Survey: \revtwo{Eliciting envisioned expressions of intended AI functionality}}
\label{sec:survey}

\revtwo{Here we report on our survey that elicited envisioned expressions of intended AI functionality by asking for delegation messages. The survey focused on how users would express their writing intent to access AI-provided functions and thus allows us to characterise the envisioned input and output of such functions.} 

\revtwo{For context,} we were originally motivated to collect samples to build an intent recognition system for conversational AI in late 2020. Now, this data provides rare insight into how people ``prompted before knowing about prompting'', \revtwo{which we later contrast with their actual behaviour with our prototypes (before and after the ``ChatGPT moment''). This contrast gave rise to the thinking behind Functional Flexibility.} %

The survey \revtwo{was also designed to} understand which tools people prefer for editing text documents and how they communicate and collaborate with others. This combination of elicited text samples and understanding the context in collaborative workflows provided the foundation for the design and implementation of our first prototype. Here, we focus on reporting the analysis of \revtwo{expressions of intended AI functionality}, and provide the results on tool use, communication, and collaboration behavior, and the complete questionnaires in our project repository: \url{https://osf.io/qh46n}.%

\subsection{Survey design}

We designed the survey to cover the collaborative writing workflow as comprehensively as possible. We researched examples from industry and academia to ask about realistic, typical tasks that people might be interested in delegating to an AI for text editing (overview in \cref{tab:survey-design}).

\begin{table*}[ht]
\caption{An overview of potential tasks users could delegate to AI. These tasks correspond to specific AI capabilities and were derived from examples from research and industry. Our survey elicited written examples for delegating these tasks by asking participants to write down how they would instruct an AI assistant in a text editor environment to perform them.}
\Description{This table shows an overview of potential tasks users could delegate to an AI. In the table, we list these tasks as AI capabilities. In total, we derived eighteen capabilities from related work and industry examples, and described them briefly in the table.}
\label{tab:survey-design}
\small
\begin{tblr}{
    rowsep = 3pt,
    colspec = {X[2,l]X[3,l]X[l]}
}
\toprule
\textbf{AI Capabilities (Tasks)} & \textbf{Description} &\textbf{Found in} \\ \hline
Revise & Revise a section of text (i.e. change or correct the text). & \cite{ter_hoeve_conversations_2020} \\
Summarize & Summarize a section of text. & \cite{ter_hoeve_conversations_2020, gehrmann_visual_2019} \\
Format  & Change a text's formatting. & \cite{ter_hoeve_conversations_2020} \\
Shorten & Shorten a section of text. & \cite{gehrmann_visual_2019, filippova_sentence_2015}\\
Extend, Elaborate & Extend/Elaborate on a sentence or paragraph that is currently rather short. & Industry: \cite{quillbot} \\
Highlight  & Highlight a part of the text. & \cite{ter_hoeve_conversations_2020} \\
Cut  & Cut a section of text that is redundant or unnecessary. & \cite{ter_hoeve_conversations_2020} \\
Generate: Complete sentence & Complete an already started sentence. & Industry: \cite{quillbot} \\
Generate: Draft paragraph from keywords & Generate a complete paragraph from an inline list of keywords. & \cite{clark_creative_2018, yang_sketching_2019} \\
Generate: Inspire/suggest & Generate a draft for a given topic. & \cite{clark_creative_2018} \\
Find and add fact and/or reference & Find fact and add reference. & \cite{yang_sketching_2019, erkens_coordination_2005} \\
Find non-textual material (e.g. photos, illustrations) & Find images or graphics that fit to the theme of the text. & \cite{koch_may_2019, koch_imagesense_2020} \\
Translate & Translate a paragraph to the given language. & \cite{radford_language_2019} \\
Question answering (e.g. domain question) & Answer a question by the user in the context of the text. & \cite{radford_language_2019} \\
Send & Send a review request to another user. & \cite{ter_hoeve_conversations_2020, birnholtz_write_2013} \\
Ideation & Make suggestions on an outline. & \cite{lu_inkplanner_2019} \\
Create outline/structure/writing plan & Suggest a text outline. & \cite{lu_inkplanner_2019, yang_sketching_2019} \\
Collaboration & AI facilitates collaboration between human writers (e.g. resolve conflicts, highlight changes, notify others) & \cite{erkens_coordination_2005} \\
\bottomrule
\end{tblr}
\end{table*}

\subsection{Survey structure}

We structured the survey into six sections: workflow, collaboration, writing stages, coordination, AI capabilities, and demographics. Brief descriptions of these sections can be found in \cref{tab:survey-structure}.

\begin{table*}[ht]
\caption{The structure of our survey.}
\Description{This table shows the structure of our formative survey. We separated the survey into six sections. The table shows the topic of each section and a brief description.}
\label{tab:survey-structure}
\begin{tblr}{X[1,r]X[9,l]}
\toprule
\textbf{Section} & \textbf{Topic \& Description} \\ \hline
1 & {\textbf{Workflow with text documents} \\
  General writing experience, tool preferences} \\
2 & {\textbf{Collaborative workflow} \\
  Communication preferences, tools for collaboration} \\
3 & {\textbf{Writing stages (Outlining, Drafting, Revising)} \\
  Asking for up to five phrases that participants would like to be able to tell an AI assistant to support them during the respective writing stage.} \\
4 & {\textbf{Coordinating work on a text} \\
  Asking for up to five phrases that participants would like to be able to tell an AI assistant to support them with coordinating work on a text (questions, task assignments, etc.).} \\
5 & {\textbf{AI capabilities} \\
  Rating of usefulness. Asking to suggest phrasings for giving a task to AI. Up to three example sentences for each AI capability, see \Cref{tab:survey-design}.}\\
6 & {\textbf{Demographics} \\
  Demographics, such as age and occupation} \\
 \bottomrule
\end{tblr}
\end{table*}

\subsection{Participants}

We consider data from 121 participants (62 female, 54 male, 4 preferred not to say, 1 other). In total, we recruited 122 participants via Prolific\footnote{\url{https://prolific.co}} but removed the data of one participant due to a mismatch with the pre-screening requirements (language: English). People's age ranged from 18 to 70 years, with a median of 32 years. Self-rating of English language proficiency revealed that 101 were native speakers, 15 were proficient, 1 had advanced knowledge, and 4 had intermediate knowledge. All participants were compensated with £\,8.8 per hour.

\begin{figure*}[t]
    \centering
    \subfloat[\centering Editor Mockup]{{\includegraphics[width=0.4\textwidth]{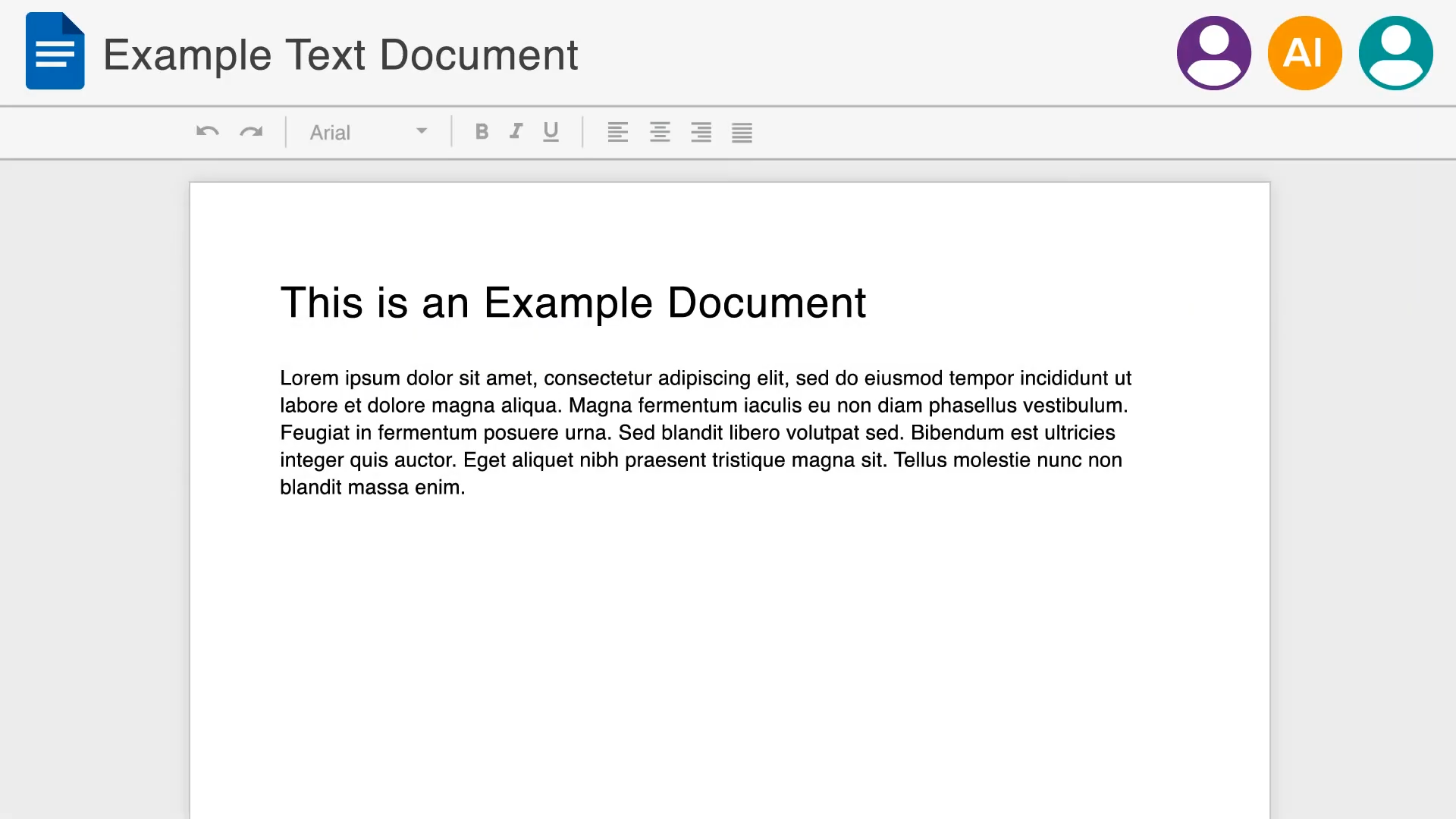} }}
    \qquad
    \subfloat[\centering Comment Mockup]{{\includegraphics[width=0.4\textwidth]{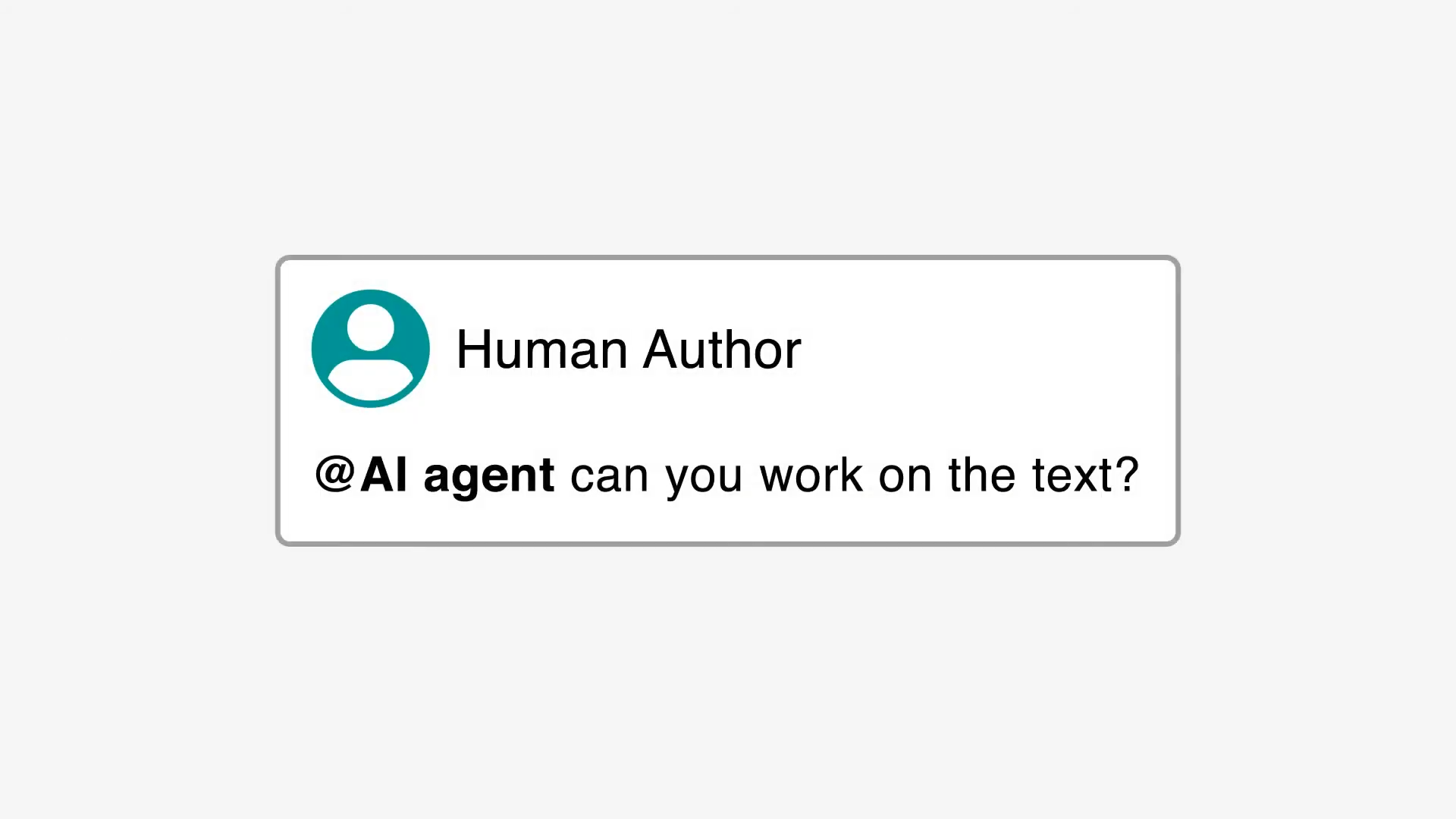} }}
    \caption{Two mockup screenshots we presented to survey participants to demonstrate how a text editor could look like that allows for delegating tasks to AI. (a) Displays the AI appearing on the document like a user (top right icons), while (b) displays a UI element with a written task delegation to AI. We intentionally decoupled the visual representation of the task delegation from the editor to put focus on the process of delegating the task, not on a specific UI design.}
    \Description{This figure shows two screenshots we presented to survey participants to demonstrate how an AI editing tool with AI capabilities could potentially look like. On the left is a mockup of a text editor that displays an AI icon to visualize the presence of the AI next to collaborators. On the right is a mockup of a card element displaying a comment delegating the task to an AI that has the capability to carry it out: ``@AI agent can you work on the text?''.}
    \label{fig:survey-explainer}
\end{figure*}

\subsection{Coding of \revtwo{envisioned expressions of intended AI functionality}}

As a central part of our survey, we asked participants to provide phrases they would like to tell an AI agent in the context of a text document editor, for the writing stages \textit{outlining}, \textit{drafting}, \textit{revising}, and \textit{coordinating}, see \cref{fig:survey-explainer}. Moreover, we elicited phrases to give certain tasks to the AI, such as revising or extending a text. In total, we collected phrases for 18 AI capabilities in this way (\cref{tab:survey-design}). %

To understand the elicited phrases in detail, we coded them for various categories, such as goals, types, and implied AI roles.
We inductively coded all phrases with a team of four coders and followed a three-step process. We first proposed initial categories and codes, and discussed them with the goal of finding about eight to twelve categories that could potentially cover the responses across all questions. This led to an initial codebook, which presented an overview of all categories, with descriptions and a list of codes. Furthermore, it contained concrete example decisions/reasonings for each category. Second, each coder independently coded the first 25 rows of each category based on the initial codebook. Coders were allowed to suggest new codes for existing categories while coding. We extensively discussed these first codings in multiple meetings to align our coding decisions and find consensus across coders. Third, we coded all samples. Each coder coded the responses for four to five AI capabilities and for one writing stage. Afterwards, our team performed a round-robin check of each others' codings. Finally, the lead coder streamlined all codings and flagged low-quality samples.

\subsection{Coding results}
We coded a total of 1447 phrases that participants would use with AI in the specific writing stages and to coordinate working on a text, as well as 3206 phrases for the specific AI capabilities. In this section, we focus on reporting on the coded categories ``Style'', ``Goals / Motivation'', and ``Co-Creative Role''. %
Tables with details about these codings %
can be found in \cref{appendix}.

\begin{table*}[ht]
\caption{Overview of our coding of elicited task delegations in the four writing stages: outlining, drafting, revising, and coordinating. Around 90\% of elicited task delegations for AI were valid (we removed samples that were flagged as low quality by the coders). The average text length was approximately 6 words per task delegation. Reading ease was computed as the mean Flesch Reading Ease Score: Delegations scored ``fairly easy'' in the coordinating stage and ``plain English'' in the others.}
\Description{This table shows an overview of our coding of elicited task delegations in the four writing stages: outlining, drafting, revising, and coordinating. It reports on the writing stage, samples, removed samples, valid samples, statistics on text length, and reading ease scores.}
\label{tab:writing-stages-general-stats}

\begin{tblr}{lrrrlr}
\toprule
\textbf{Writing Stage} & \SetCell[r=1]{l}{\textbf{Samples}} & \SetCell[r=1]{l}{\textbf{Removed}} & \SetCell[r=1]{l}{\textbf{Valid}} & \SetCell[r=1]{l}{\textbf{Text Length (words)}} & \SetCell[r=1]{l}{\textbf{Reading Ease}} \\ \hline
Outlining     & 444                         & 67                              & 377 (84.91\%)             & M=5.59, SD=3.14, min=1, max=21 & 64.84                            \\
Drafting      & 407                         & 50                              & 357 (87.71\%)             & M=5.30, SD=3.31, min=1, max=19 & 62.66                            \\
Revising      & 403                         & 41                              & 362 (89.83\%)             & M=5.32, SD=3.17, min=1, max=18 & 62.71                            \\
Coordinating  & 390                         & 39                              & 351 (90.00\%)             & M=6.96, SD=3.77, min=1, max=22 & 71.58          \\
\bottomrule
\end{tblr}
\end{table*}

\begin{table*}[ht]
\caption{Overview of our codings of elicited task delegation inputs for using specific AI capabilities. We covered 18 AI capabilities but focused on the three capabilities in bold, namely summarize, extend, and translate. 83.60\% to 95.59\% of inputs were valid, e.g. we removed samples with a low quality. The average text length was between four and five words. The mean Flesch Reading Ease Score showed delegations to AI for summarizing a text as ``difficult to read'', and delegations to AI for extending and translating as ``plain English``.}
\Description{This table shows an overview of our coding of elicited task delegation inputs for using specific AI capabilities. It reports for each of the 18 AI capabilities the total samples, removed samples, valid samples, statistics on text length, and reading ease scores.}
\label{tab:ai-skills-general-stats}
\small
\begin{tblr}{lrrrlr}
\toprule
\textbf{AI Capability} & \SetCell[r=1]{l}{\textbf{Samples}} & \SetCell[r=1]{l}{\textbf{Removed}} & \SetCell[r=1]{l}{\textbf{Valid}} & \SetCell[r=1]{l}{\textbf{Text Length (words)}} & \SetCell[r=1]{l}{\textbf{Reading Ease}} \\ \hline
Revise                 & 204                         & 12                          & 192 (94.12\%)          & M=4.14, SD=3.10, min=1, max=18          & 62.60                            \\
\textbf{Summarize}     & \textbf{189}                & \textbf{31}                 & \textbf{158 (83.60\%)} & \textbf{M=4.96, SD=6.28, min=1, max=54} & \textbf{56.61}                   \\
Format                 & 223                         & 14                          & 209 (93.72\%)          & M=4.32, SD=2.70, min=1, max=16          & 61.20                            \\
Shorten                & 206                         & 33                          & 173 (83.98\%)          & M=4.68, SD=3.95, min=1, max=40          & 64.38                            \\
\textbf{Extend}        & \textbf{194}                & \textbf{26}                 & \textbf{168 (86.60\%)} & \textbf{M=4.54, SD=2.91, min=1, max=18} & \textbf{66.55}                   \\
Highlight              & 198                         & 12                          & 186 (93.94\%)          & M=4.41, SD=4.41, min=1, max=54          & 64.24                            \\
Cut                    & 206                         & 14                          & 192 (93.20\%)          & M=4.32, SD=2.81, min=1, max=18          & 65.70                            \\
Complete               & 191                         & 27                          & 164 (85.86\%)          & M=4.52, SD=3.32, min=1, max=22          & 55.59                            \\
Draft                  & 198                         & 25                          & 173 (87.37\%)          & M=5.49, SD=3.28, min=1, max=19          & 71.77                            \\
Inspire                & 193                         & 19                          & 174 (90.16\%)          & M=5.50, SD=4.24, min=1, max=34          & 64.61                            \\
Add Reference          & 205                         & 12                          & 193 (94.15\%)          & M=4.56, SD=3.13, min=1, max=17          & 49.30                            \\
Find Non-Text Material & 216                         & 34                          & 182 (84.26\%)          & M=5.03, SD=2.68, min=1, max=17          & 65.72                            \\
\textbf{Translate}     & \textbf{204}                & \textbf{9}                  & \textbf{195 (95.59\%)} & \textbf{M=4.67, SD=3.20, min=1, max=21} & \textbf{62.92}                   \\
Question Answering     & 210                         & 23                          & 187 (89.05\%)          & M=5.03, SD=2.83, min=1, max=17          & 77.73                            \\
Send                   & 206                         & 7                           & 199 (96.60\%)          & M=4.93, SD=3.12, min=1, max=18          & 80.67                            \\
Ideation               & 185                         & 43                          & 142 (76.76\%)          & M=4.96, SD=2.93, min=1, max=14          & 68.29                            \\
Create Outline         & 197                         & 33                          & 164 (83.25\%)          & M=4.68, SD=2.72, min=1, max=15          & 62.09                            \\
Collaboration          & 204                         & 49                          & 155 (75.98\%)          & M=5.86, SD=3.30, min=1, max=19          & 56.47                        \\
\bottomrule
\end{tblr}
\end{table*}

\subsubsection{\revtwo{Task perspective: Eliciting expressions of intended AI functionality based on task stage/needs}}

\cref{tab:writing-stages-general-stats} shows descriptive statistics. The length of the phrases varied between 1 and 22 words. The mean length per stage was between 5.32 and 5.59 words; only Coordinating had 6.96. The coded ``goals'' (\cref{tab:writing-stages-goals}) showed that participants aimed for inspiration and learning in the outlining stage. For example, participants wrote phrases such as ``AI agent please suggest a document structure'' and ``look up this heading in wikipedia''. In the drafting and revising stage, they focused on efficiency and improving text, such as ``Check content for punctuation'', ``Check for flow and ease of reading'' when drafting, and ``Correct spelling errors'' and ``Reword this text for me'' when revising. The predominant goal in the Coordinating stage is coordination, and less focused on efficiency, for example, ``Divide tasks evenly amongst contributors'' and ``Keep track of to-do tasks''.

As shown in \cref{tab:writing-stages-styles}, the coding of input ``styles'' revealed that imperative inputs (e.g. ``Suggest additions'') appeared the most, while almost a quarter of inputs were questions (e.g. ``Can you give some feedback on this?''). %
Regarding ``roles'' (\cref{tab:writing-stages-roles}), participants looked primarily for AI as an assistant, in particular for coordinating, for example, with phrases like ``Send the document out for review or approval''. For outlining, they also referred to the AI in a co-author role, with phrases such as ``Can you add another section to the body?''.

\subsubsection{\revtwo{AI capability perspective: Eliciting expressions of intended AI functionality for given AI capabilities}}

As shown in \cref{tab:ai-skills-general-stats}, phrases for giving specific tasks to an AI had between 1 and 54 words, with a mean length per AI capability between 4.54 and 4.96 words. We focus on the AI capabilities \textit{summarize}, \textit{translate}, and \textit{extend} (bold in the table), since we deemed those technically feasible for our prototype at the time. The motivations for delegating a summarization task to the AI were primarily efficiency and improving the text, such as ``Summarise highlighted section'' and ``Help and summarize the meaning of this text, to ensure it is reflected in the first paragraph''. In contrast, translations were motivated purely by efficiency (i.e. to replace the need for manual translation), %
as the coding of ``goals'' revealed in \cref{tab:ai-skills-goals}. When delegating text extension, participants were also motivated to gain inspiration, for example, with phrases like ``I need more content on this''. Coding of ``styles'', as shown in \cref{tab:ai-skills-general-style}, revealed that participants largely used an imperative style (i.e. giving instructions). Finally, their phrases imply that the AI acts in the role of a co-author (see \cref{tab:ai-skills-roles}), for example ``Write a longer alternative.'' %

\begin{table*}[ht]
\caption{Key findings from our survey (N=121) \revtwo{with respect to the concept of Functional Flexibility.}}
\Description{This table presents the key findings from our survey. It shows each aspect and a description of the key findings.}
\label{tab:takeaway-survey}
\begin{tblr}{p{4cm}X}
\toprule
\textbf{Key Finding}                                  & \textbf{Details}      \\ \midrule
\revtwo{Elcitited concrete expressions of intent-driven functions} & \revision{The survey collected concrete examples of how users would express their intent to access intent-driven writing functions in the functional space offered by an LLM.}\\
\revtwo{Expressions varied across task stage and AI capability} & \revtwo{The elicited expressions for accessing AI functions revealed various assumed roles and styles across task stages and AI capability, reflecting varying assumptions about needs and expected outcomes\revthreeflo{, which is consistent with the view that flexibility is a key characteristic of such AI functions in the writing context}.} \\
\revtwo{Desired functions were mostly expressed in full sentences} & Across writing stages, people wrote between five and seven words \revtwo{to access an AI function}. \revtwo{Looking at specific expected capabilities,} delegations for the tasks summarize, extend, and translate had a length between four and five words. \\
\revtwo{Desired functions assumed controllable scope} & \revtwo{When expressing desired AI functions, participants pointed the AI to scopes such as words, sentences, sections, and the whole document, revealing scope as an intuitively assumed input parameter.} \\
\bottomrule
\end{tblr}
\end{table*}

\subsection{Survey summary}

An overview of the key findings is given in \cref{tab:takeaway-survey}.
\revthreeflo{In summary, we interpreted the survey analysis regarding both the \textit{input} and \textit{output} side of Functional Flexibility:}
\revision{Regarding the input, by collecting samples across writing stages, we captured a high-level picture of conceptual functions instructed to AI. Complementary, the phrases for delegating a certain task revealed insights on a detailed level per AI function.} 
\revtwo{It is worth highlighting that people mostly envisioned writing full sentences here, with a} mean length of five and seven words across writing stages. Looking at the AI capabilities summarize, extend, and translate, the phrases had a mean length between four and five words. \revtwo{We will return to this when contrasting the survey and study results.}

\revtwo{Regarding the output,} the intents expressed in the elicited phrases reveal users' expectations about AI-generated text \revision{(i.e. the output of conceptual functions)}. 
\revtwo{Together, these reveal concrete mappings and ``parameters'' for flexibly accessing AI functions. Crucially,} users assumed to interact with certain text parts as input to conceptual functions, \revtwo{which in turn} imply different scopes \revtwo{for AI} -- for example, words, sentences, sections, the whole text. 

\revision{Overall, this analysis supported us in understanding \revtwo{the input and output space for flexibly accessing AI functions, that is,} how users express and anchor intents prompted to AI, and the conceptual functions they are aiming to fulfil.}

\section{Prototype 1 - Conversational UI}

\revtwo{In the process of our research (\cref{fig:framing}), the survey also informed our first prototype:}
When asked for delegations for specific capabilities, participants predominantly assumed an AI co-author.
For our first prototype, this aspect motivated us to design a user interface in which the assisting AI appears akin to a ``co-author'' (e.g. via presence indicators and comments known from human-human collaboration). 
\revtwo{Moreover, the result of implied scopes -- combined with the reported use of comment features when writing with other people -- supported our design choice to integrate access to AI functions via comments on user-selected (i.e. scoped) text.}

We combined \revtwo{these} findings from our survey with a design session in our research group to inform how to integrate AI capabilities into a text editor. %
Note that at the time of designing this prototype (2021), the conversational abilities of language models were far more limited than today, and conversational UI patterns like in ChatGPT or a ``Copilot sidebar'' did not yet exist (cf.~\cite{yuan_wordcraft_2022} in 2022). %

\subsection{Design concept}
\label{sec:prototype1-final-concept}

\revtwoflo{In our research group design session, we generated UI design ideas, and followed the approach of integrating AI into interactive comments, i.e. letting users define AI functions in interactive comments.}

While designing these features, we identified spatial, temporal, and informational design dimensions as cornerstones for designing a conversational AI. These design dimensions impact positioning, delays, and communication with the AI.

\begin{figure*}[t]
    \centering
    \includegraphics[width=0.9\textwidth]{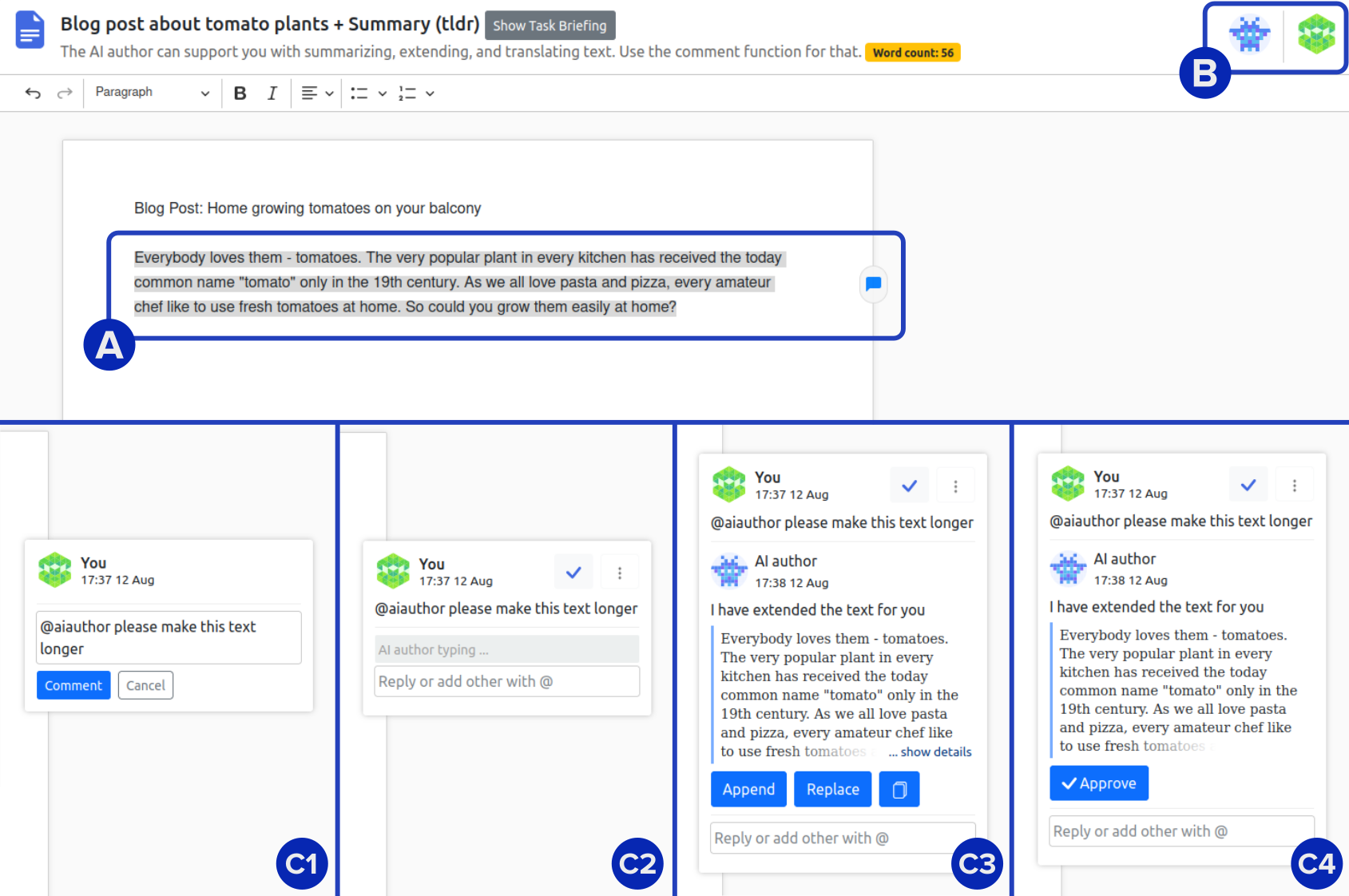}
    \caption{Screenshots of our first prototype with a conversational UI. The top half of the figure shows the editor. We designed the editor to look similar to existing online text editing tools. Part A shows a user-selected text, with a balloon icon next to it on the right for adding an interactive comment. Part B shows that both the user and the AI are ``present'' in the document. The bottom half of the figure shows the procedure of delegating a task to the AI via an interactive comment \revtwoflo{to access an AI function}. C1 shows a comment to extend the text. C2 shows an indicator that the AI is carrying out the task (``typing''). In the top right of the comment card, users can approve the comment, or use the dropdown to edit or delete the comment. C3 shows the AI's suggestion, three options to accept the text (append, replace, copy to clipboard). Furthermore, users can view more details and potentially write a reply. However, our system did not support multi-turn conversations. C4 shows the comment card after the users have clicked on ``append'' and the suggestion has been appended in the main text -- users can then approve this result, which closes (removes) the comment.}
    \Description{This figure shows screenshots of our first prototype with a conversational UI. The top half of the figure shows the editor, which looks similar to existing online text editing tools. The bottom half shows the process of adding an interactive comment, delegating a task to the AI, the AI responding, and accepting an AI suggestion.}
        \label{fig:prototype_1}
\end{figure*}

The final design consists of four major parts and is shown in \cref{fig:prototype_1}. This figure shows (A) the text editor that allows for writing with AI. We integrated the AI similar to a human collaborator. (B) The AI is ``embodied'' in the top right of the screen with a user icon. This user bar signals the availability of the AI. The bottom half of the figure zooms in on interactive comments next to the page view. The comments (C1) allow for conversations \revtwoflo{for delegating tasks as AI functions}, (C2) addressing users/AI (@aiauthor), (C3) presenting suggestions, and (C4) interacting with the suggestions, such as accepting and rejecting them. The comments displayed a reply input field, \revtwoflo{to give users the chance to refine an AI function,} yet multi-turn conversations were not supported. Moreover, the comment interface provided three different options to work with the suggestion (appending it to the selected text, replacing the selected text, and copying it to the clipboard). It also offered a detail view for the suggestions, see \cref{fig:prototype_1_diffview}.

Using interactive comments as the central interaction element for AI has the benefit of providing a visual reference to the annotated parts of the document while being visually decoupled from it. Comments are also particularly used for task-oriented communication \cite{birnholtz_write_2013} by users. With such annotations, the interaction is less intrusive, is user-initiated, and lets writers stay as close as possible to their usual writing workflow \cite{clark_creative_2018}. Having the comment cards in the UI also served as a space for the AI to suggest changes \cite{wang_why_2017} and for the user to track the history of such generated snippets. Moreover, we designed the AI agent to behave akin to a human in its use of the comments. For example, an indicator showed when the AI ``typed'' an answer, and response times were randomly extended by 10 to 19 seconds.

\subsection{Technical implementation}

We prototyped the frontend with React\footnote{https://react.dev/}, CKEditor 5\footnote{https://ckeditor.com/}, and Bootstrap\footnote{https://getbootstrap.com/}. We implemented a CKEditor Plugin for marking and referring to text. The intent recognition \revtwoflo{for delegating tasks to AI} was based on the RASA framework\footnote{https://rasa.com/}. %
Concretely, we trained a Dual Intent and Entity Transformer classifier with RASA in the default configuration (epochs set to 100) and used the phrases collected from our survey as the training dataset. We used this intent recognition to call the fitting AI \revtwoflo{functionality} from the backend, given a user's comment text. This AI \revtwoflo{functionality} was provided through separate REST APIs based on FastAPI\footnote{https://fastapi.tiangolo.com/}. The generative models were T5 for summarizing text, Opus-MT for translating text, and GPT-Neo for extending text. %

\section{Study 1: \revtwoflo{Observing expressions of intended AI functionality in actual use with a conversational UI}}
\label{sec:study1}

\revtwo{Following up on the survey, we investigated how people express intended AI functionality in actual use, concretely,} how people integrate AI into their workflow \revtwoflo{by expressing intended AI functionality} in interactive comments to delegate tasks for extending, translating, and summarizing. The study was \revision{conducted online with our prototype in the role of a technology probe~\cite{Feng2025canvil, Hutchinson2003techprobe}}. Combining qualitative and quantitative methods, we gained a comprehensive picture of users' interactions and experiences. %

\subsection{Task} 

We designed the task to engage participants to \revtwo{express intents for AI functionality} for different aspects of writing, \revtwo{following the findings of the survey}. Concretely, participants wrote a blog post about a neutral topic, growing tomato plants at home, in English. The topic might further require a short research phase, depending on individual familiarity with it. Participants were only allowed to use certain online resources. We provided them with a list of hyperlinks to existing guides and blog posts about tomato plants in German. Thus, this task provided a realistic opportunity for \revtwoflo{expressing intents for AI translation}. %
Participants had to finish the blog post within 30 minutes -- thus, the task provided an incentive for \revtwoflo{expressing intents for the extending AI function}. Furthermore, people were instructed to add a summary of three sentences about their blog post -- thus, the task also included an opportunity \revtwoflo{for expressing intents for the summary AI function}. Participants finished the task with a mean time of 40.01 minutes (SD=7.97, min=28.81, max=55.61).

\subsection{Apparatus}

Participants used our conversational UI prototype implemented as a web app, as described in \cref{sec:prototype1-final-concept}. They used a web browser on their own computer to take part in the study and joined an online call with the experimenter.

\subsection{Participants}

Ten people participated in our study (4 female, 6 male). Their age ranged from 27 to 36 years, with a median of 30. Five participants self-reported to speak English well, four very well, and one fairly well. Self-reports for German language proficiency showed nine native speakers and one person to speak German very well. All participants were recruited via social media and compensated with 12\,€ per hour.

\subsection{Procedure}

\begin{figure*}[t]
    \centering
    \includegraphics[width=0.85\textwidth]{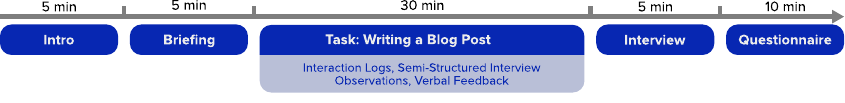}
    \caption{The procedure of our online experiment took approximately one hour. After reading an intro text and giving informed consent, users read a task briefing that also included a video to introduce the features. Participants could ask for clarification. Then, their task was to write a blog post about growing tomatoes at home. We recorded the interaction logs, observations, verbal feedback, and the shared screen. This was followed by a semi-structured interview. In a final step, participants filled in a questionnaire with subjective ratings and questions about socio-demographics.}
    \Description{This figure visualizes the procedure of our experiment. It is a timeline, displaying the parts of the procedure: from left to right: 5 minutes Intro, 5 minutes Briefing, 30 minutes Task of writing a blog post with a semi-structured interview, observations, and logging of interactions, 5 minutes interview end, and a final questionnaire.}
    \label{fig:procedure}
\end{figure*}

Participants joined a video call with the experimenter who provided them a hyperlink to a web application that guided them through the experiment. The experiment's procedure is depicted in \cref{fig:procedure} and took about one hour.

In the first step, the web application presented an introduction about the study and data protection regulations, and participants had to give their informed consent. This step was followed by the task briefing (\cref{apx:task-briefing}). People read the instructions and watched an explainer video. They could ask the experimenter for clarification. In the third step, participants used the conversational UI prototype to complete the task. They shared the screen (web browser window) via the video call software, which we recorded. The experimenter asked questions of the semi-structured interview throughout the session, depending on user interaction with the tool and completed the interview after the task had been done. For the final step, the screen sharing and recordings were stopped, and participants filled in a questionnaire. It covered socio-demographics, language proficiency, and Likert items on usability aspects, including the SUS questionnaire \cite{brooke_sus_1996}. %

\subsection{Results}

We analyzed the quantitative log data with a focus on how users interacted with AI through the conversational UI \revtwoflo{to express intended AI functionality}. We complement this analysis with qualitative findings. %

\subsubsection{Task delegation \revtwoflo{for accessing intended AI functionality}}

We analyzed how participants delegated writing tasks to the AI through the conversational UI. \revtwoflo{This task delegation reflects how people express their intent to access AI functionality.}
They added a total of 139 comments, of which 131 were sent to the AI, while eight were removed without writing an instruction for AI. Each participant posted a mean of 13.1 comments (SD=4.41, min=5, max=17) with a mean length of 1.79 words (SD=0.89, min=1, max=3.8). In four cases, participants tried to combine AI skills by using the word ``and'' in the comment (e.g. ``translate \textit{and} extend''). Writing a comment took a mean time of 8.93 seconds (SD=12.52, min=2, max=119).

\begin{figure}
    \centering
    \includegraphics[width=0.6\linewidth]{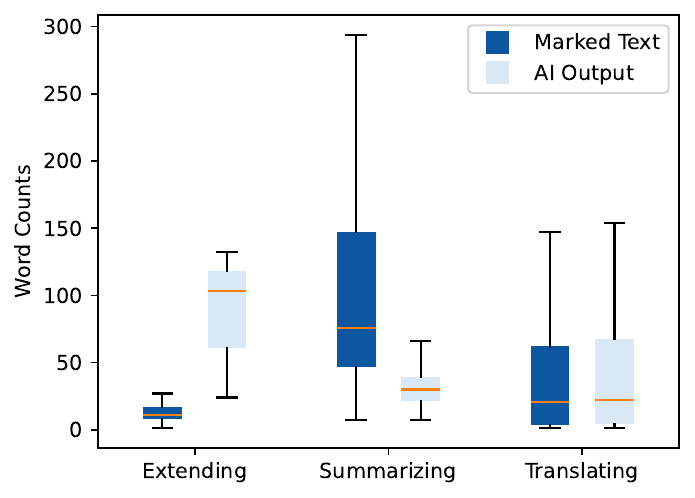}
    \caption{Lengths of AI text generations and the lengths of the users' selected text, measured as number of words. This visualization shows how users employ AI functions to modify text with varying length characteristics: \textit{Extending} is applied on short text selections, and the AI then generates text that is several times longer. With \textit{summarizing}, input text length varies the most, and output is much shorter. The ratio for \textit{translating} is close to one.}
    \label{fig:cui_marked_text_generations}
    \Description{This figure shows paired box plots of the lengths of marked text and AI suggestions for each AI capability, namely generation, summarization, and translation.}
\end{figure}

We counted 128 AI replies. Three comments were removed by the participants before the AI reply arrived. AI replies had a mean length of 52.96 words (SD=44.88, min=1, max=201). \textit{Extensions} were the longest with a mean length of 90.21 words (SD=34.20, min=24, max=312), followed by \textit{translations} with a mean length of 43.95 words (SD=52.74, min=1, max=201). \textit{Summaries} were the shortest with a mean length of 30.95 words (SD=12.88, min=7, max=66). \cref{fig:cui_marked_text_generations} shows these lengths of AI responses next to the lengths of the users' corresponding text selections. This figure reveals that users apply the AI capabilities to transform text in different ways. When extending the text, users select short snippets, and the AI generates text that is several times longer. In contrast, with summaries, longer input text is shortened considerably by the AI. With translations (German-English), the ratio is close to one.

The analysis of comment editing behavior showed that participants input a mean of 13.93 keystrokes per comment (SD=10.5, min=2, max=71). This excludes editing actions (backspace, clear, cut, delete). Analyzing these separately, participants corrected their input with a mean of 2.5 editing keystrokes per comment (SD=2.13, min=1, max=12). Eight participants deleted parts of their input with a mean sequence length of 2.72 editing keystrokes per affected comment (SD=0.75, min=2, max=4). These sequential edits happened in 18 comments.

\subsubsection{Acceptance and rejections}

We analyzed acceptances and rejections of AI-generated comments\revtwoflo{, as a proxy measure on how well the AI functionalities served the user intent}. %
Users could explicitly accept these suggestions via the three buttons \textit{append}, \textit{replace}, and \textit{copy to clipboard}. We counted these interactions as acceptances. 

Furthermore, participants could delete comments, which we count as rejections. They could also resolve a comment at any time, which has more nuance: We counted resolve events as approvals of the suggestions when users chose to integrate them by appending or copying the text to the clipboard beforehand. If no such accept events happened beforehand, we counted resolve events as a negative signal, adding to the rejection count. %

\begin{figure}[t]
    \centering
    \includegraphics[width=0.5\linewidth]{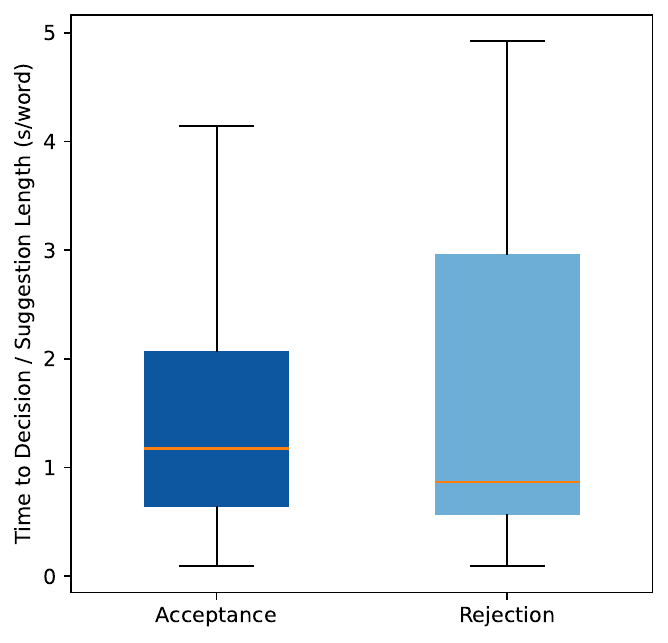}
    \caption{The time participants took to decide whether to accept or reject an AI-generated comment, normalized by the length of the generated text. This plot shows that acceptances happened faster than rejections.}\label{fig:cui_accpting_rejecting_time_to_decision}
    \Description{This figure visualizes the participants' time to arrive at a decision for accepting or rejecting an AI suggestion. The time to arrive at a decision is divided by the suggestion length (s/word).}
\end{figure}

\Cref{fig:cui_accpting_rejecting_time_to_decision} visualizes the time that participants needed to arrive at a decision to accept or reject an AI suggestion, normalized by its length. On average, participants were slightly slower to accept suggestions, but the times for rejection varied more. %

\paragraph{Acceptance:}

AI suggestions were accepted a total of 86 times, with a median normalized time to acceptance of 1.17 s/word (SD=9.45, min=0.09, max=82.5). Relative acceptance rates per AI capability revealed that translations were accepted most often with 83.33\% (40 of 48), followed by summarizations with 65.11\% (28 of 43), and extensions with 48.64\% (18 of 37).

\paragraph{Rejection:}

Users rejected a total of 35 suggestions, with a median normalized time to rejection of 0.87 s/word (SD=20.53, min=0.09, max=112). Rejections per AI capability happened most often for extensions with 48.64\% (18 of 37), followed by summarizations with 27.90\% (12 of 43), and translations with 10.42\% (5 of 48).

\subsubsection{Final Texts}

Overall, participants produced blog posts with a mean length of 386 words (SD=52.92, min=334, max=491). We computed the Flesch reading-ease score as a criterion for readability. The mean reading ease was ``fairly easy to read'' (72.04) and ranged from ``Easy to read. Conversational English for consumers.'' to ``Difficult to read.'' (SD=10.10, min= 49.93, max=84.88).

\begin{table*}[ht]
\caption{Overview of the key findings from our user study (N=10) on writing with AI on a text document using a conversational comment-like UI. \revtwoflo{These findings capture how people express/access intended AI functionality through a conversational UI.}}
\Description{This table shows an overview of the key findings from study 1. It lists aspects and a description of each key finding.}
\label{tab:takeaway-study1}
\begin{tblr}{p{4cm}X}
\toprule
\textbf{Aspect}            & \textbf{Key Finding}        \\ \hline
\revtwoflo{Command-like task delegation for accessing AI functionality}    & Participants used very short, command-like instructions for delegating tasks with a mean length of 1.79 words. \revtwo{This puts this observation from actual use in contrast to the longer prompts in the survey.}                  \\
\revtwoflo{Varying text selections for scoping AI functions} & Participants selected text depending on the desired \revtwoflo{AI functionality} (e.g. shorter selection for extending text, longer selections for summarizing text) \\
\revtwoflo{Increased acceptance of AI outputs that serve the user intent}        & Suggestions were accepted more often if they fulfilled the user's intent. \revtwo{Reaching this decision, i.e. accepting suggestions, took more time than rejecting them.} \\
\revtwoflo{Specified AI functions as an alternative to conversations}   & Participants \revtwo{expressed a need for} pre-determined AI tools, e.g. via an AI toolbar, \revision{confining the flexibility for accessing AI functions.}    \\    \bottomrule   
\end{tblr}
\end{table*}

\subsection{User perception: verbal feedback, observations, and ratings}

To get a more complete picture of the interactions, we asked participants in a post-hoc questionnaire to rate Likert statements on usability aspects. %
An overview of the ratings can be found in \cref{fig:cui_ratings_overview}, and the questionnaire is available in \cref{tab:study1-questionnaire}. 
The mean SUS score was 83.5, ranging from 67.5 to 92.5 (see \cref{fig:cui_sus_ratings}), indicating excellent usability \cite{bangor_empirical_2008}. Furthermore, the majority of participants found the AI system natural (3 agree, 4 strongly agree), see \cref{fig:cui_additional_ratings}. Also, the majority agreed that the tool made them reach their goal faster (3 agree, 5 strongly agree) and supported them to complete their task (5 agree, 3 strongly agree). %

The experimenter took notes with the approximate timestamps while participants worked on the task. These notes captured surprising or interesting observations of the participants' interactions and comments. The notes were categorized, for example, as reactions to the AI capabilities, the participants' overall editing strategy. Another researcher of our group reviewed the recordings, notes, and categories. In a brief discussion, we clarified and corrected unclear notes. 
These categorized observations and responses to the semi-structured interviews revealed \revtwo{the following aspects:}

\subsubsection{Specific use of AI and writing strategies} 
All participants used the AI \revtwoflo{functionalities} in specific ways and integrated them into their writing strategies. For example, this included translating single words (P2, P9, P10) and extending partial sentences (P3, P7, P10). Users also adapted their overall writing strategy with AI. For example, P5 started with manual writing and intended to apply AI functions later, while P4 took another approach by translating a long text from the sources and then summarizing the resulting paragraphs manually. \revtwo{Regarding the theoretical construct presented in this paper, these varied integrations supported our focus on flexibility as a fundamental characteristic of interest in conceptualising user-desired AI-provided functions.}%

\subsubsection{Conversations with AI vs. AI as tools}
Four out of the ten participants tried to converse with the AI by trying to reply to a comment\revtwoflo{, for example to refine an intended AI function.} P5 suggested further conversation could be used to switch to another AI \revtwoflo{functionality} if intent recognition failed. In contrast, three participants (P2, P3, P5) in the interview requested to put AI \revtwoflo{functions} into tools (e.g. buttons) rather than delegating tasks with natural language. \revtwo{Comments such as these imply that participants themselves thought about varying degrees of flexibility in accessing AI functionality (e.g. open conversation vs. specified buttons). This observation motivated our thinking on functional flexibility as something to shape via the UI elements, instead of simply maximizing it by (only) providing a free prompt entry UI.}

\subsubsection{Mental models of AI and interpretation of its inner workings}
For six participants, their comments revealed that they reasoned about how the AI worked internally. In particular, factors such as delay and context influenced their interpretations, as reactions show, such as by P6 ``It takes longer, that must mean something good.'', or P10 ``I thought it would be faster for shorter text'', and P7 ``I thought the AI reads the thing [part of the text] and wouldn't need keywords as input''. Some participants also attributed more general, almost human-like behavior and skills to the AI, such as P2 reacting surprised about a generation, ``cool, it's googling for me'', as well as P9 ``oh he thinks, that I have started in June, interesting''. \revtwo{Such comments suggest that the mental model can influence how users try to access AI functions (e.g. different scoping behaviors, or different assumptions about the required info in their prompts). Since UI design influences mental models~\cite{norman_design_2002}, this observation further motivated our thinking on the role of the UI in our concept of functional flexibility as shaping how users access the functional space provided by AI.}

\subsection{\revision{Conversational UI summary} \revtwoflo{and implications for the concept and the second prototype}}

\revtwo{In summary, the findings of this study (\cref{tab:takeaway-study1}) revealed several insights that shaped our conceptual thinking behind functional flexibility and functional space: Concretely, the findings \revthreeflo{supported our thinking on} flexibility, which is shaped by the UI design, as this \revthreeflo{defines} how users can access the functional space provided by an LLM.}

\revtwo{An interesting tension emerged when contrasting the expressed intents for AI functions with those from the survey:} \revision{Our prototype offered a high degree of functional flexibility in accessing AI capabilities by expressing user intent in natural language in comments to AI. However, participants accessed AI functions through} short, command-like instructions, %
\revtwo{in contrast to the more elaborate expressions envisioned in the survey.}

\revtwo{Further moving away from the idea of elaborate, open expressions of intent, some} participants explicitly requested an AI toolset \revtwo{that would avoid the need for repeatedly writing prompts.} \revtwo{In terms of our concept,} \revision{they thus asked for reduced flexibility in accessing functions provided by LLMs.} %

\revtwo{Together, these} findings %
motivated us to \revtwoflo{also explore more restricted yet targeted UIs for accessing specific AI functions, with another prototype.} %
Our second prototype thus provided AI \revtwoflo{functions} as part of a toolbar instead of a conversational UI. \revtwoflo{At the same time, we included free prompting in the design to still offer a high degree of functional flexibility, with which users are able to define own functions.} \revtwo{Thus, with the second prototype we also explored offering users a mix of UI elements with different degrees of functional flexibility.} 

\section{Study 2: \revtwoflo{Observing expressions of intended AI functionality when the UI offers multiple degrees of functional flexibility}}\label{sec:study2}

\label{sec:prototype_design_changes}
As mentioned above, the findings from our first study informed how we adapted the prototype for our second user study. %

\subsection{Changes to the UI based on design implications of study 1}

Participants in our first study wrote short commands for delegating tasks to the AI. This included single-word commands, such as ``translate'', ``extend'', or even just ``ext''. %
Furthermore, participants requested to use the AI functions through buttons. For example, P5 suggested to ``put the functions into buttons or floating tooltips'', P3 would rather ``use the functions without the comments'', and P2 mentioned ``AI capabilities could have been implemented into the context menu''. Compared to the elicited task delegations in our survey (\cref{sec:survey}), users thus ``minimized'' their input for delegating tasks to AI during actual interaction in a text editor. Concretely, the mean length of inputs in our first study was 1.79 words, while in the survey it was between 4 and 5 words. Thus, both qualitative and quantitative results pointed us into the same direction: People preferred efficient commands over open delegation prompts.
Interestingly, in retrospection, participants' behavior and wishes prior to exposure to ChatGPT thus went against the recent product trend of adding very open prompting \revision{as central} features in the UI, which assume elaborate \revision{prompting} (e.g. chatbot assistants in sidebars).

Based on these findings, we decided to replace the conversational UI with a toolbar UI. This toolbar offered the three \revtwoflo{pre-specified AI} functions \textit{extend}, \textit{translate}, and \textit{summarize} as buttons. Furthermore, it offered a button that opens a text entry field for prompt input. At the time (early 2023), such prompting had become widely available via systems such as ChatGPT. \revtwo{Thus, the prototype} offered both fixed AI functions \revtwoflo{(limited flexibility)} and open prompts \revtwoflo{(high flexibility)}. 

As a part of introducing the toolbar UI, we also changed the text highlighting/markers. In our new prototype, we used different colors, one for each AI function (extend, summarize, translate, free prompt). Furthermore, we \revision{changed} the detail view to a column layout that displayed the user's selected text on the left and the system's suggestion on the right, to ease comparison (\Cref{fig:prototype_2_diffview}).

\subsection{Changes to the system responses}
\begin{figure*}[t]
    \centering
    \includegraphics[width=0.9\textwidth]{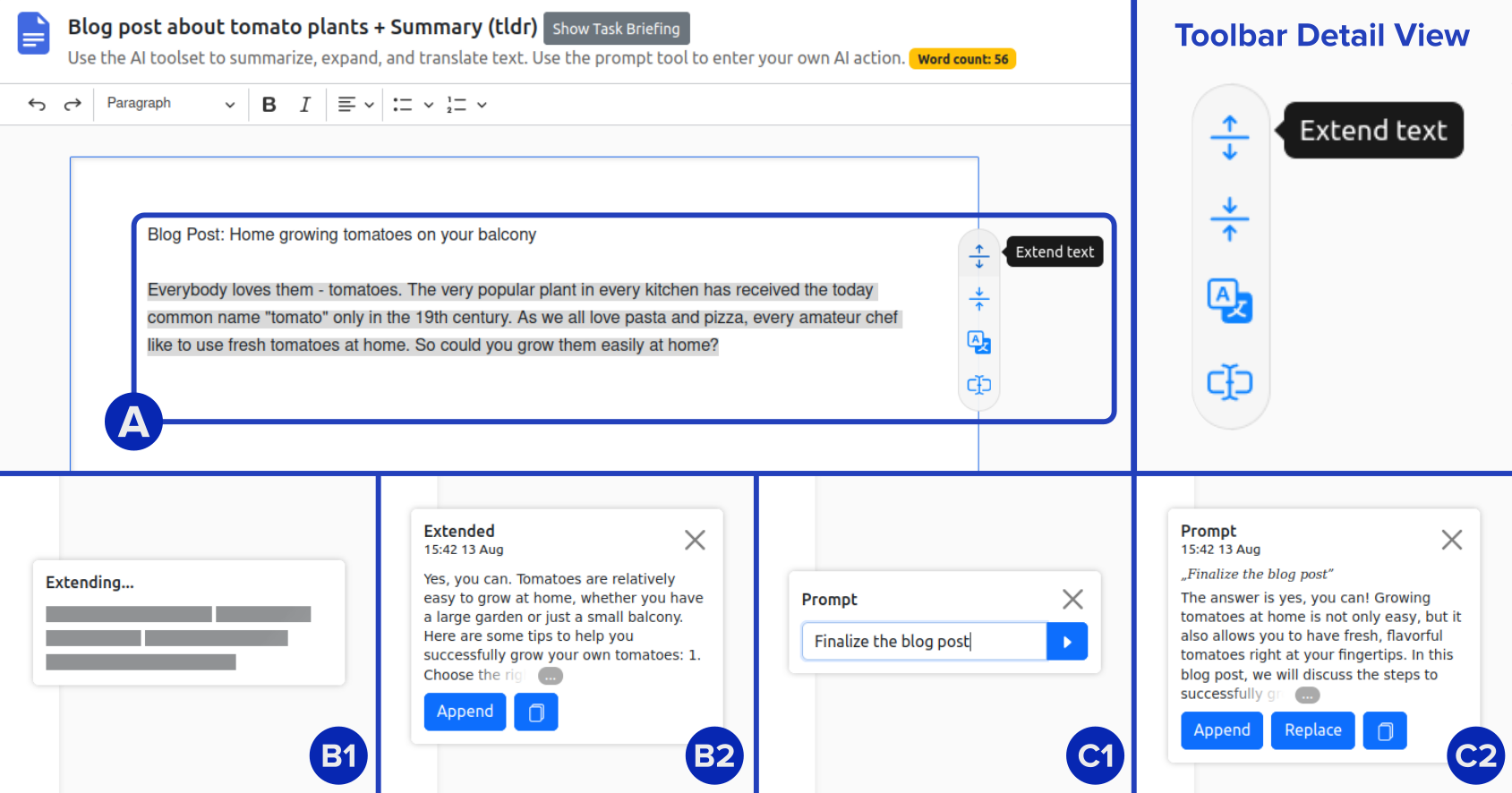}
    \caption{Screenshots of our second prototype with an AI toolbar. The top half shows the editor, similar to existing online text editor tools. Part (A) shows how users could select text, which brings up the AI toolbar floating next to it on the right. This toolbar provided buttons for extending, summarizing, translating, and prompting own functions. The bottom half shows the procedure of applying an AI function from the toolbar. (B1) shows a loading indicator after a user has clicked on the extend text function. (B2) shows the AI's suggestion, with buttons for viewing the complete suggestion (``...''), appending it, and copying it to the clipboard. Replacing (cf.~\cref{fig:prototype_1}) was not possible for ``extend'' since it would have overwritten the source text. (C1) shows the UI for prompting own functions. (C2) shows the AI's response to this.}
    \Description{This figure shows screenshots of our alternative prototype with an AI toolbar. The top half shows the editing tool that looks similar to existing online text editing tools. The bottom half shows the processes of using a pre-determined AI function from the toolbar, and the prompting of user-tailored functions.}
    \label{fig:prototype-2}
\end{figure*}

We changed the system response as well. In the first prototype, we used comment-card elements floating at the right edge of the document page to include the conversational UI. We kept these cards for the second prototype but with reduced interactivity. Comments were completely removed, such that the cards only display the AI response. This new design is shown in \cref{fig:prototype-2}. The figure shows (A) the floating toolbar UI offering pre-determined AI functions and a fourth button to access free prompting. (B1) shows a loading indicator that represents the AI executing the task, and (B2) shows the AI's suggestion, an icon to show the details, and options for accepting the suggestion. For \textit{extend}, the generated text could only be appended or copied to the clipboard. That is, we removed the ``replace'' button since this would have overwritten the user's original text, which does not match the user expectation for \textit{extend}. (C1) shows the interface for free prompting, and (C2) shows the corresponding AI response. The cards could be closed (removed) through the cross icon in the top right.

\subsection{Implementation changes}

We switched the model for extending text from GPT-Neo to GPT3, and used this model also for executing custom prompts. The models for summarizing and translating text remained the same.

\subsection{Study design \revtwo{and participants}}

The design for our second user study remained the same as in study 1 (\cref{sec:study1}). The only change was the prototype. %

We recruited 12 participants (5 female, 7 male). Their median age was 28.5 years (range: 20-36). Five self-reported their English language proficiency as very well, another five as well, one as fairly well, and another one as poor. Eleven people were German native speakers. One rated to speak German well. All participants were recruited via social media and compensated with 12\,€ per hour.

\subsection{Results}

Similar to the first user study, participants finished the writing task with a mean time of 39.57 minutes (SD=9.01, min=22.85, max=51.48).

\subsubsection{Data cleaning}

In total, participants used the AI tools 243 times. Of these, 35 events had an empty AI response or no response at all.\footnote{Note that at the time of conducting the study, the OpenAI web API was under high load and randomly timed out.} We thus excluded these 35 events and considered the remaining 208 usages of AI tools for our analysis.

\subsubsection{Use of pre-determined AI tools and free prompting}

\begin{figure}[t]
    \centering
    \includegraphics[width=0.5\linewidth]{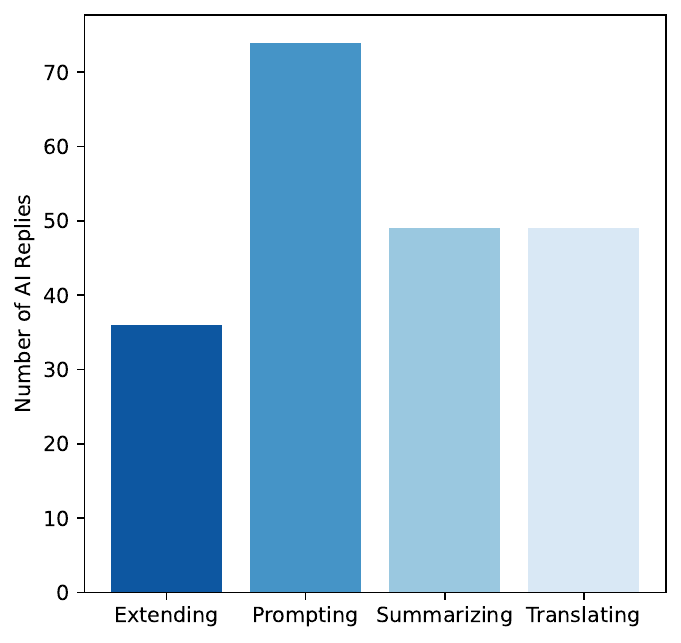}
    \caption{Use of the AI tools in our second user study. Free prompting was used most often, followed by pre-determined tools for summarizing and translating. Extending text was used the least.}
    \Description{This figure shows a bar chart of the AI tools usage in our second user study. It shows bars for the pre-determined tools, generation, summarization, and translation, as well as prompts.}
    \label{fig:tools_number_ai_replies}
\end{figure}

On average, participants used AI tools 17.33 times (SD=7.0, min=7, max=30). They prompted a total of 74 own functions (35.57\% of AI uses), while the summaries and translations were used 49 times each (23.56\%), and text extension was used 36 times (17.31\%), as shown in \cref{fig:tools_number_ai_replies}. Only one participant tried to combine AI functions by using the word ``and'' in the prompt, namely ``integrate both sentences \textit{and} shorten thereby''.

Each participant prompted a mean of 6.17 own functions (SD=2.82, min=1, max=11) with a mean length of 5.59 words (SD=2.93, min=1.5, max=10.33). AI responses on these free prompted functions had a mean length of 59.82 words (SD=45.92, min=2, max=205), overall AI responses were shorter with a mean of 56.39 words (SD=45.43, min=2, max=205). 

\begin{figure}[t]
    \centering
    \includegraphics[width=0.75\linewidth]{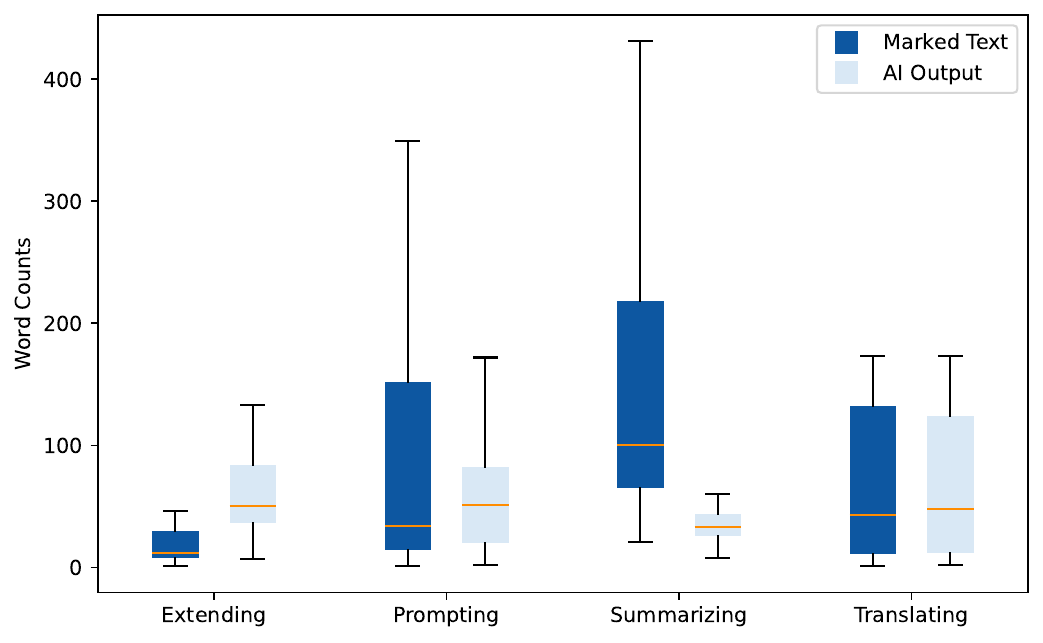}
    \caption{Paired lengths of AI generations and user-selected text, measured as number of words. With free prompting of functions, AI output resulted in shorter text compared to the original, selected text. The characteristics of the other AI tools remained similar to our first user study (cf.~\cref{fig:cui_marked_text_generations}.}
    \label{fig:tools_marked_text_generations}
    \Description{This figure shows paired box plots of text lengths of marked text and AI suggestions for each pre-determined AI tool (generation, summarization, and translation) as well as prompts.}
\end{figure}

\cref{fig:tools_marked_text_generations} shows the characteristics of the input and output lengths. For extend, summarize, and translate, these are similar to the results in study 1 (see \cref{fig:cui_marked_text_generations}). %
Freely prompted functions make the text longer on average, and the lengths of the resulting responses vary less compared to the input.

We also analyzed the editing behavior when prompting own functions. Participants input a mean of 37.88 keystrokes per prompt (SD=27.62, min=2, max=126). This excludes editing actions/keys such as backspace, clear, cut, and delete. 
These editing actions/keys had a mean of 4.76 keystrokes per prompt (SD=4.3, min=1, max=17). %
Nine participants pressed editing keys in sequence. They sequentially deleted input with a mean of 3.49 delete-keystrokes per prompt (SD=1.57, min=2, max=7). This sequential editing occurred 21 times. %

\subsubsection{Acceptance and rejection}

\begin{figure}[t]
    \centering
    \includegraphics[width=0.5\linewidth]{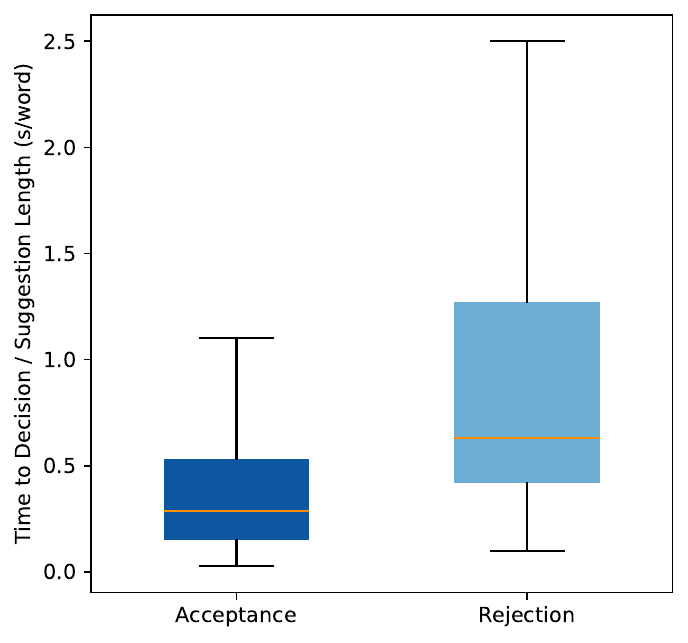}
    \caption{Visualization of the time that users spent on deciding to accept or reject AI-generated text. The times are normalized by the lengths of the generated texts. This plot shows that acceptances happened faster than rejections. Rejections might have taken people more time since they might be cognitively more demanding.}
\label{fig:tools_accpting_rejecting_time_to_decision}
    \Description{This figure shows a box plot of the time participants needed to arrive at a decision for accepting or rejecting AI suggestions. The time is normalized by the suggestion length (s/word).}
\end{figure}

As in the first study, we analyzed acceptances and rejections of AI-generated text. %
Similar to the first version of our prototype, users could accept an AI generation by appending, replacing, and copying it to the clipboard. We counted these interactions as acceptances. We counted rejections when closing cards without any such acceptance interaction. \cref{fig:tools_accpting_rejecting_time_to_decision} shows the decision times for acceptances and rejections, normalized by the length of the suggestions. As in study 1, participants accepted suggestions faster than rejecting them.

\paragraph{Acceptance:}

A total of 137 AI generations were accepted. The median normalized time to acceptance was 0.29 s/word (SD=1.21, min=0.03, max=12.67). 
Separated by functions, translations were accepted the most with 87.76\% (43 of 49), followed by prompted functions with 67.57\% (50 of 74), extensions with 58.33\% (21 of 36), and summarizations with 46.94\% (23 of 49).

\paragraph{Rejection:}

Participants rejected a total of 56 AI suggestions. The median normalized time to rejection was 0.63 s/word (SD=2.62, min=0.09, max=61.5). People most often rejected summaries with 48.98\% (24 of 49), followed by prompts with 27.03\% (20 of 74), extensions with 22.22\% (8 of 36), and translations with 8.16\% (4 of 49).

\subsubsection{Final Texts}

Participants created blog posts with a mean length of 339.83 words (SD=29.59, min=304, max=410). The Flesch reading-ease score showed a mean readability of ``Plain English'' (68.75) and ranged from ``Easy to read'' to ``Fairly difficult to read.'' (SD=7.97, min=56.76, max=81.43).

\subsection{User perception: verbal feedback, observations, and ratings}

As in study 1, we analyzed the qualitative data to learn about participants' perceptions when interacting with the prototype. They rated Likert items on usability via the a SUS questionnaire and additional items on topics. An overview of these ratings can be found in \cref{fig:tools_ratings_overview}. The questions are listed in \cref{tab:study2-questionnaire}.

The mean SUS score was 87.29 (SD=6.94, min=77.50, max=97.5), see \cref{fig:tools_sus_ratings}. This score reflects an ``excellent usability'' \cite{bangor_empirical_2008}, and is above the score of our first prototype. \cref{fig:tools_additional_ratings} shows the additional ratings, which revealed that almost everyone found the interaction with the system natural (7 agree, 4 strongly agree). Furthermore, all participants found that the system supported them in their task (3 agree, 9 strongly agree). Ratings on efficiency were similarly positive (4 agree, 7 strongly agree). In contrast to our first study, we further asked about control and authorship. A majority of participants agreed to be in control of editing the text during the task (5 agree, 4 strongly agree). Ratings on feeling like the author of the text were ambiguous (4 agree, 1 disagree, 3 strongly disagree).

Since tools such as ChatGPT had become popular shortly prior to conducting our second user study, our questionnaire asked if participants had prior experience with intelligent writing tools. All indeed had such experiences: Eleven had experienced auto-correction and eight had used ChatGPT for writing.

We recorded each session and took notes of observations as in the first study. Data processing also stayed the same.

Based on the participants' feedback and our observations, we identified four themes. %

\subsubsection{Users were unsure about prompting and wished for inspiration.} 
Eight of twelve participants mentioned that they were unsure about prompting. For example, P4 was not familiar with it. Others struggled to come up with ideas for prompting -- as commented by P5: ``intuitively, I have not got an idea, extend already exists''. Responses in the semi-structured interview revealed that they would have wished for examples of prompts, such as P7: ``I wish for a list of examples, I was not sure if I was free about text input.'' Similar, P9 noticed ``I used only primitive commands'' and expected that a set of examples would allow them ``to use it [prompting] more flexible''. We provided participants with an initial example in the briefing and they were free to input and try out their own prompts. However, it seems this freedom hindered them at first, as also reflected on by P12. A reason for this challenge might be the creative and open nature of such prompts. P6, for example, first liked the translate feature the most but then favored the prompting of own functions since ``it is the most creative''. However, P4 perceived this as a demand (``you have got to be very creative'') and P9 reflected: ``I ask myself, how I can prompt more creatively''. Guidance for prompting might support lowering such creative barriers; as P12 put it ``give examples for prompting to lead the user''.

\revthreeflo{We interpret these observations as pointing to a relationship between the degree of functional flexibility – and how a UI makes the functional space accessible to users – and the users’ experience. Concretely, they suggest a tradeoff: high functional flexibility can be more demanding, as users have to put effort into identifying and expressing their intents.}

\subsubsection{Users adapted existing functions to improve quality and control over the output} We built our system by combining several models to provide the AI functions (translations via Opus-MT, summaries via T5, and GPT3 for extending text and free prompts). These models showed varying output quality. All participants gave feedback on how they perceived this. %
For example, P1 found the extend output too vague: ``not the story I wanted to go for [...] America is too general [as the origin of tomato plants].'' On the other hand, P3 found it useful: ``I am surprised about the length, actually it is good''. For the AI summaries, participants remarked on lower quality. P8 found that the ``summary is unsatisfying'' and P10 ``that is not a good summary''. 

When expectations about output quality were not met by the pre-determined AI tools, participants prompted replacement functions. For example, six participants prompted their own summary functions. Some of them used the free prompting to further control their replacement function. For instance, P3 said that they are ``prompting for summary since then I can be more specific''. Similarly, P9 remarked ``the length of the summary should be adjustable'', followed by specifying the length of the summary in an own prompted summary function. %

\revtwo{\revthreeflo{These observations suggest that UI elements with various degrees of functional flexibility may inform each other’s use.} For example, while open prompting was difficult (see above), experiencing pre-specified AI-provided functionality via the toolbar gave people concrete prompting ideas for improvement -- \revthreeflo{in other words, low flexibility UI elements appeared to provide guidance for using high flexibility ones.} Thus, regarding our theoretical construct, these results supported our thinking on the role of the UI design overall (and designers) in shaping how users can access the functional space.}

\label{sec:serial-combination-study-2}
\subsubsection{Chaining multiple AI tools} Participants chained the different tools to reach their goals as part of their writing strategy. P5 and P6, for example, first translated German text into English and then extended the text. P8 prompted a function and then summarized the rather long result to shorten it. P9 combined various functions to modify parts of a text. This specific use of AI tools shows that participants utilize AI tools as ``atomic'' functions and combine them to reach an intended goal, \revtwo{again supporting the idea that the UI elements -- with varying degrees of functional flexibility -- in combination shape how users access the functional space provided by an LLM.}

\subsubsection{Mental model about AI and interpretations of the context} %
Even with the toolbar UI, participants %
reasoned about the system's inner workings, similar to our first prototype. %
For instance, P1 asked ``is only selected context of interest?'' and P4 said it ``[...] adds more detail. I found that also in the sources. Now, I know where that came from.'' Similarly, P5 interpreted AI tools to involve external sources: ``oh, he did research for me, it is a good support for research''. This comment also reveals anthropomorphism (``he''), although we had removed the ``embodied'' representation from study 1 (e.g. no user icon for the AI). Reactions by P6 and P9 also showed this: When the API took a few moments to respond, P6 said, for example, ``well, it seems he has to think about that.'' %

\revtwo{These comments suggest that varying degrees of functional flexibility -- while changing how users can access the functional space -- do not necessarily change how users think about the system's inner workings since the output characteristics are still those of generative AI (e.g. whether text generation was triggered via a prompt or button, it still leads to the system adding generated text). This supports the focus in our concept on how users \textit{access} the functional space (rather than, say, what the system output is) when using it as a thinking tool across specific UI designs (see \cref{sec:funcflex_thinking_tool}).} %

\begin{figure}[t]
    \centering
    \includegraphics[width=0.5\linewidth]{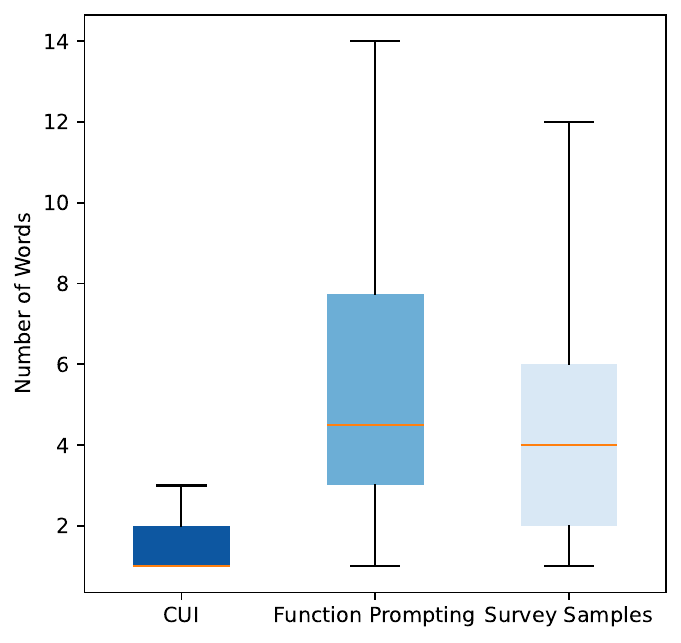}
    \caption{In study two, participants used the free prompting tool with longer prompts compared to the text that participants in study one entered with the comment feature. The prompts reached lengths similar to those of the task delegation samples envisioned by participants in our survey.}
    \label{fig:comparison_prompt_lenth}
    \Description{This figure shows a box plot of the text length inputs had for delegating a task when using our conversational UI, prompting user-tailored functions with our toolbar UI, and from our formative survey.}
\end{figure}

\begin{table*}[t]
\caption{Key findings from our second user study (N=12) on working with AI on a text document by using a toolbar UI that offers pre-determined AI tools as well as free prompting.}
\Description{This table shows an overview of our key findings from study 2. It lists the aspects and descriptions of the key findings.}
\label{tab:takeaway-study2}
\begin{tblr}{>{\raggedright}p{14em}X}
\toprule
\textbf{Aspect}            & \textbf{Key Finding}            \\ \midrule
\revtwo{UI designs beyond maximum flexibility: offering varying degrees of functional flexibility} & \revision{Our second prototype offered two ways of using AI, varying the functional flexibility (AI toolset: limited flexibility, free prompting: maximum flexibility). \revthreeflo{Our observations suggest that the functional flexibility accounted for in UI designs shapes how people integrate AI into their workflow.}} \\
\revtwo{Acessing the functional space with high flexibility (i.e. writing prompts) can be challenging but pre-defined (i.e. low flexibility) tools provide indications of the functional space} & \revtwo{Participants were unsure about how to prompt in general. At the same time, they used prompting to extend, refine, improve AI functions originally provided in the UI via dedicated buttons, which gave them concrete ideas of possible functions.}\\
People \revtwo{strategically} combined AI tools and free prompting & Pre-determined AI tools (extend, summarize, translate) accounted in sum for 64.42\% of all triggered AI functions. Participants prompted their own functions in 35.57\% of all uses of AI. These prompts had a mean length of 5.59 words, \revtwo{bringing the level of elaboration (prompt length) back to the lengths elicited in the survey -- for those intents that participants chose to express with the full flexibility UI, instead of clicking a toolbar button.} \\
\bottomrule
\end{tblr}
\end{table*}

\subsection{\revision{AI toolset \& free prompting summary} \revtwoflo{ -- how users switch between different degrees of functional flexibility as implemented in tools and free prompting}}

\revision{\revthreeflo{With this second prototype, we offered UI elements with different degrees of functional flexibility side by side for accessing AI functions.} The pre-determined tools offered limited flexibility while the free prompting offered a maximum of flexibility.} See \cref{tab:takeaway-study2} for a summary. 

\revtwo{These findings \revthreeflo{served as} further \revthreeflo{evidence to guide our theory-building, and} strengthened and refined our conceptual perspective \revthreeflo{on the} theoretical construct of functional flexibility and functional space: Concretely, a maximum of flexibility (i.e. free prompting) can be demanding, motivating UIs that offer varying degrees of flexibility via different UI features, which in turn can impact each others' use, since pre-defined AI functions can serve as indications of the functional space (i.e. of what's possible with high flexibility elements) and users also strategically switch between the two and combine them, to access the functional space provided by an LLM.}

\revtwo{To highlight some concrete observations in this regard:} Most participants were unsure about how to use free prompting at first, although 75\% had experience with tools such as ChatGPT. Nevertheless, after some exploration, they applied self-prompted functions, for example, \revtwo{to create custom functions or} improved versions of existing AI tools, \revtwo{which thus could be seen as low flexibility UI elements providing guidance for high flexibility ones}. They used AI tools as ``atomic'' functions and combined and chained them to reach their goals.

Their prompts entered via the free prompting box were longer than the conversation prompts entered in study 1, as shown in \cref{fig:comparison_prompt_lenth}. \revthreeflo{This observation is consistent with the view that free prompting of functions offers users more functional flexibility than a limited set of functions (summarize, translate, extend), at the cost of more effort in specifying their intent. As the two studies differed in more than the UI, we treat this as an observation that informed our construct, not as an effect of functional flexibility.} %
Interestingly, the length of these freely prompted functions was similar to the length of what people had envisioned to enter as task delegations in our survey. \revtwo{One interpretation of this is that when given both low and high flexibility options, users can go with the low ones if they do not see the need to express more specific intentions, but when they do choose the flexible option (i.e. free prompting) they then invest more into expressing their intended functionality.}

\revthreeflo{Overall, our observations in the second study suggest that the functional flexibility accounted for in UI design can shape how people integrate AI into their workflow. Together with our survey and first study, these observations served as theory-building evidence for conceptualizing functional space and functional flexibility as a theoretical construct.}

\section{How user interfaces shape access to the functional space of generative AI}
\label{sec:functional_space}

\begin{table*}[t]
\caption{An overview of the core aspects of our theoretical construct on how user interfaces shape access to the functional space of generative AI.}
\Description{This table provides an overview of the key aspects of our theoretical construct. It lists the most important aspects and our theoretical approaches.}
\label{tab:takeaway-space}
\begin{tblr}{lX}
\toprule
\textbf{Aspect}           & \textbf{Theoretical approach}   \\ \hline
Functional Space & LLMs offer great functional flexibility which promises users an unlimited space of output generations that fulfill desired functions. Thus, an LLM can be considered as a functional space. Users instruct LLMs with text inputs from a similarly unlimited input space to access these functions. \\
Role of User Interfaces  & UIs define how users can express their intents and to which degree they can access the LLM's functional space. Concretely, UIs allow diving into the functional space at different locations and limit the reachable subspace.                                                                        \\
Key Insight   & %
The key mechanism by which ``interaction with LLMs'' can be construed is accessing a desired target function through suitable UIs.
\\
\bottomrule
\end{tblr}
\end{table*}

\revision{Here, we} bring together the reported findings of our empirical investigations, \revthreeflo{as guiding our theory-building,} with contributions from related work to introduce \revision{the theoretical construct of \textit{functional space and functional flexibility}, } for thinking about text-based interaction with generative AI: Its key concept is the \textit{functional space} offered by LLMs to users, an unlimited space of output generations that \revision{fulfill conceptual} functions. Such \textit{functions} serve a user's goal. They might come with rather clear expectations, such as translating and summarizing a text, but might also be more fuzzy, such as inspiring and reflecting. \textit{The role of the UI} is to determine how users can express their intents (e.g. click a summary button vs. enter a prompt ``summarize this'') and to which degree they can access their desired \textit{target function} within this functional space. Through this conceptual lens, we describe how different UIs make AI functions accessible in different ways. An overview of these core aspects is shown in \cref{tab:takeaway-space}.

Following Generative Theories of Interaction \cite{beaudouin-lafon_generative_2021}, we have compiled a set of questions that can be asked to interpret the analytical, critical, and constructive power of functional flexibility as realized in a system, see \cref{tab:generative-theory-questions}. These questions help HCI researchers evaluate and refine AI systems, guiding the design of interfaces to offer the right degree (or mix) of functional flexibility.

In the remainder of this section, we take an analytical approach to examine our own applications \revision{as well as a recent example from related work} through the lens of functional flexibility to address the question of ``Can existing systems be deconstructed in terms of functional flexibility?'' (cf. ~\cite{beaudouin-lafon_generative_2021}).
The discussion section then builds on this analysis, offering a critical and constructive use of functional flexibility, and \revision{further} outlooks \revision{on its applications} (\cref{sec:discussion}).

\begin{table*}[t]
\caption{Questions that can be asked to interpret the analytical, critical, and constructive power of functional flexibility implemented in a system \cite{beaudouin-lafon_generative_2021}. \revthreeflo{Q3 - Q8 focus on how UIs make the functional space of generative AI accessible to users, where functions are often only determined once users express their intents, going beyond the traditionally developer-specified reifications of commands in Instrumental Interaction~\cite{beaudouin-lafon_instrumental_2000}.}}
\Description{This table provides a numbered list of questions to take an analytical, critical, or constructive perspective on interpreting a system in terms of functional flexibility.}
\label{tab:generative-theory-questions}
\begin{tblr}{colspec={llX}, rowsep=1pt}
\toprule
\textbf{Perspective} & \textbf{ID} & \textbf{Questions} \\
\midrule
Analytical   & Q1  & What UI elements are available with which degree of functional flexibility? \\
             & Q2  & What mix of functional flexibility is offered by the UI / interaction design overall? \\
             & \revthreeflo{Q3} & \revthreeflo{Which functions does a UI element make accessible, and how many?} \\
             & \revthreeflo{Q4} & \revthreeflo{Who defines a function, and when (system builder at design time vs.\ user at use time)?} \\
             & \revthreeflo{Q5} & \revthreeflo{How do users identify and express a desired function, and how can the UI support this?} \\
             & \revthreeflo{Q6} & \revthreeflo{How do UI elements with different degrees of functional flexibility influence each other's use?} \\
             & \revthreeflo{Q7} & \revthreeflo{How can the set of directly accessible functions evolve with use?} \\
             & \revthreeflo{Q8} & \revthreeflo{How should a UI account for underspecified, non-deterministic functions?} \\
\midrule
Critical     & Q9  & How does the degree of functional flexibility influence the interaction with the system? \\
             & Q10 & How does the offered degree of functional flexibility improve the system? \\
             & Q11 & Does the system/UI offer the right degree of functional flexibility to the user for reaching/identifying their goals? \\
             & Q12 & What problems does the degree of functional flexibility offered by the system cause? \\
             & Q13 & Can users reach/identify their goals with the degree of functional flexibility offered by the system? \\
\midrule
Constructive & Q14 & Can functional flexibility inspire new ideas when designing a new system? (Inspired by \cite{mackay_interaction_2025}) \\
             & Q15 & How can users be supported to reach their goal? \\
             & Q16 & How can we help users identify their goal? \\
             & Q17 & Which UI elements could be added to a design to increase/decrease functional flexibility? \\
\bottomrule
\end{tblr}
\end{table*}

\subsection{Grounding the theoretical construct of functional flexibility}

\revision{In contrast to smaller, more specific models built for a certain task, the development of language models gave rise to general-purpose LLMs~\cite{brown_language_2020, radford_language_2019} that serve a multitude of tasks. Such tasks fulfill conceptual \textit{functions} that are based on user intents. Examples of intent-based functions in the context of text editing include paraphrasing, summarizing, or drafting a text \cite{mysore_prototypical_2025}. Thus, general-purpose models offer a large \textit{functional space}, that promises users almost unlimited possibilities of output generations. Users explore these possibilities by instructing LLMs from an similarly unlimited input space of written text. This view is related to other recent conceptual HCI work: \citet{reynolds_prompt_2021} described interactions with LLMs as locating an already learned task (i.e. locating an already existing function). Finding the right words to locate such tasks is termed ``abstraction matching'' by \citet{liu_what_2023} and identified as the ``gulf of envisioning'' by \citet{subramonyam_bridging_2024} (i.e. describing a function). To relate such (envisioned) functions to their use via a UI, we can also follow the theory of instrumental interaction and reification~\cite{beaudouin-lafon_instrumental_2000, Beaudouinlafon2000reification}.} %

\revthreeflo{Instrumental Interaction by \citet{beaudouin-lafon_instrumental_2000} leverages the manipulation of graphical UI objects known from direct manipulation, by reifying abstract commands in interface tools for users to interact with}: In this light, an LLM can be seen both (1) as a \revthreeflo{(part of a) way of realising the functionality of} a tool, for example, for \revthreeflo{summarising} a text document, but also (2) as an object of interest \revthreeflo{(that needs reifications to be manipulated)}, for example when \revthreeflo{developing UIs for} engineering prompts for an LLM. In that sense, Instrumental Interaction and Functional Flexibility both build conceptual bridges from abstract functions to UI representations. %
\revthreeflo{However, the concepts differ in level of abstraction and concern. Functional Flexibility is not limited to UIs that use direct manipulation or are tool-like in the sense of Instrumental Interaction: It instead sits on a higher level of abstraction across different kinds of UIs (e.g. incl. conversational UIs), since its key concern is how different UIs make the functional space of AI models accessible to users in different ways. This includes those UIs, where users themselves need to define functionality (e.g. by asking for it in an open prompt); and also UIs, where users can create their own reifications (e.g. by storing a prompt as a button for later re-use). This is in contrast to the original reification concept~\cite{Beaudouinlafon2000reification}, which also described user-defined tradeoffs between simplicity and control, yet as predefined by developers (i.e. the ``Layers'' in ~\cite{Beaudouinlafon2000reification}), not by end-users. Here, Functional Flexibility is concerned exactly with this new flexibility, which --  enabled by LLMs -- UIs now often grant/delegate to end-users. That said, based on our insights here, we indeed call for what could be seen as a reminder of the strengths of Instrumental Interactions in light of this new flexibility: Considering Functional Flexibility in design means considering which degree(s) of flexibility to give in a UI; and answering this may very well result in prespecified settings for users to choose from, akin to the ``Layers'' in the original reification paper. To make this analytical role concrete, a set of related analytical questions enabled by Functional Flexibility can be found in \Cref{tab:generative-theory-questions}. These questions complement established properties for comparing interaction techniques, such as those of Instrumental Interaction~\cite{beaudouin-lafon_instrumental_2000}, which characterize instruments that reify deterministic commands.} In addition, we extend prior perspectives on AI tools \cite{riche_ai-instruments_2025} by examining how different UIs shape users' access to such AI functions.

\revision{We combine these ways of thinking about interactions with LLMs, based on our reflections across \revthreeflo{the presented studies}, to extract the key insight that interacting with LLMs is about \textit{accessing a desired target function} through suitable UIs. For example, prompt-based UIs expect from the user the capability to express their intents effectively in natural language. These UIs present a free text field for entering prompts, as implemented in our second prototype, see \cref{fig:prototype-2} C1. The freedom to define any user intent, and the need to define it well \cite{subramonyam_bridging_2024}, makes it difficult for users to access the right function within the large functional space offered by LLMs. In contrast, toolbar UIs offer pre-determined AI functions to users and thereby ease access at the cost of restricting the flexibility \revtwoflo{as implemented in our second prototype}, see \cref{fig:prototype-2} A.}

\revision{In summary, w}e capture \revision{our concept} in the new theoretical construct of \textit{functional space and functional flexibility}. This provides a conceptual lens that highlights the role of user interfaces for LLMs and how they shape users' access to functions provided by LLMs. Our theoretical construct puts fresh emphasis on the role of UIs as the essential connection between the users' intents and the functional space of LLMs. In this way, UIs define both (1) how users can express their intents and (2) to which degree users can access the LLM's space of functions. Together, this determines the \textit{degree of functional flexibility} made available to users. 

\subsection{\revtwoflo{Formalizing functional flexibility as an exploratory system of conceptual functions}}

\revision{We now discuss another angle \revtwoflo{on the} concept: Our perspective on the flexibility offered through UIs to access conceptual functions in the space offered by LLMs, is similar to identifying creative concepts in a conceptual space, as introduced by \citet{wiggins_preliminary_2006}. Thus, we \revtwoflo{retrospectively define} our concept formally based on Wiggins' formalization \cite{wiggins_framework_2019}.} \revtwo{Our motivation with this is to provide further grounding and connection of the concept to prior work -- now in terms of formalization -- beyond the conceptual grounding and connections discussed above. }

\revision{First, we define key aspects of this formalization: }

\begin{itemize}
\item \revision{\(U_{LLM}\): The universe. A functional space containing all possible pieces of text.}
\item \revision{\([[.]]\): An interpreter for selecting text pieces from \(U_{LLM}\). This is the API of an LLM.}
\item  \revision{\(〈〈., ., .〉〉\): Another interpreter for traversing U and its subsets. This is the user.}
\item  \revision{\(R_{P}\): Rules defining a function. These are instructions to the LLM\revtwoflo{, e.g. a prompt and the context. With prompt inputs (cf. Study 2), people have maximum freedom in defining their own functions, while with specified AI tools, the functions are determined by the system builder.}}
\item  \revision{\(T_{UI}\): Rules for traversing the functional space. This is the UI, as it determines how a user explores the space of an LLM, \revtwoflo{i.e. with conversational UIs as in Study 1, people reply on AI outputs to iterate the output of the AI function captured by the expressed intent. With the UI in Study 2, people referred to self expressed functions in one try, or jumped directly to a subspace by using a button that captured a predefined AI function.}}
\item  \revision{\(E_{I}\): Rules for evaluating a function. Users evaluate if a function's output matches their intent. \revtwoflo{This is reflected, e.g., in Study 1 where participants accepted AI suggestions more often if they fulfilled their intent, and in Study 2 where participants prompted improved versions of the predefined functions when an AI tool output did not match their expectation.}}
\item  \revision{\(F =[[R_{P}]](U_{LLM})\): A particular functional space. This is a subspace of the almost infinite functional space.}
\end{itemize}

\revision{Then, we define the process of exploring the functional space of an LLM as follows:}
\revision{
\begin{equation} \label{eqn:funcflex}
f_{out} =〈〈R_{P}, T_{UI}, E_{I}〉〉([[R_{P}]](U_{LLM}))
\end{equation}
}

\revision{Through passing instructions \(R_{P}\) to an LLM \(U_{LLM}\) (directly through prompts, or indirectly through pre-determined instructions and interaction), a user accesses the functional space \(F\). The user \(〈〈., ., .〉〉\) explores this functional space through the UI \(T_{UI}\) and evaluates if the output matches their intent \(E_{I}\). The output results in functions \(f_{out}\), respectively a set of text generations fulfilling conceptual functions. }

\revision{Thus, our theoretical construct of functional flexibility can be expressed as:}

\revision{\begin{equation} \label{eqn:funcflex_subset}
〈U_{LLM}, [[.]], 〈〈., ., .〉〉, R_{P}, T_{UI}, E_{I}〉
\end{equation}
an exploratory system \cite{wiggins_framework_2019} of conceptual functions offered by LLMs to users.}

\subsection{Deconstructing how UIs make the functional space of generative AI accessible to users}

\begin{figure*}[t]
    \centering
    \includegraphics[width=1\textwidth]{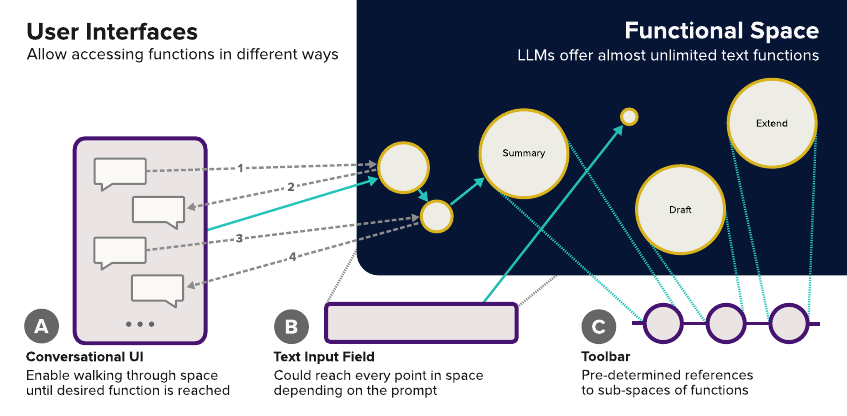}
    \caption{High-level illustration of how different degrees of functional flexibility arise from UIs that make the functional space of generative AI accessible to users: The dark plane visualizes the functional space provided by an LLM. To keep this figure generalizable, the space is given no explicit dimensions. Subspaces within this space represent different text functions that users desire to access, i.e. their target functions, such as ``summary''. Each point represents a specific function (e.g. a specific way to summarize). The illustrated UIs are examples of actual UIs that allow users to dive into the functional space at different locations and limit the reachable subspaces: (A) With a conversational UI, users typically ``walk'' through the space via messages until a desired function is reached in conversation. (B) With a text input field for free prompting, users could directly reach every point in the space, depending on the prompt. (C) A toolbar offers pre-determined references to subspaces of functions. Overall, these UIs vary in the amount of involved context that is passed to the LLM, the possibility to prompt instructions, and subsequent user interactions on the LLM's responses. Our theoretical perspective emphasizes the role of the UIs as the essential connection between user intent and the functional space of LLMs. UIs define both (1) how users can express their intents and (2) to which degree they can access the LLM's space of functions.}
    \Description{This figure visualizes our theoretical construct of how different degrees of functional flexibility arise from UIs that make the functional space of generative AI accessible to users. The illustration is separated into two parts. An area displaying three example UIs, namely a conversational UI, a text input field for free prompting, and a toolbar UI. These examples refer to another area, the functional space offered by LLMs. The UI examples point to discs within this functional space. Each disc symbolizes a text function.}
    \label{fig:space}
\end{figure*}

\Cref{fig:space} visually summarizes how UIs make the functional space of generative AI accessible. The functional space provided by an LLM is visualized as a plane without specific dimensions to serve generalizability. Hypothetically, dimensions could be \revision{labelled} depending on use cases, for example, sentiment~\cite{kim_cells_2023}, style~\cite{sterman_interacting_2020}, or could be dynamically derived from prompts \cite{suh_luminate_2024}. Regions within this space (subspaces) represent different text functions that a user might want the model to execute. We decided to use subspaces, not single points, to reflect that many text generation systems are non-deterministic and text transformations can be underspecified. For example, if a user thinks of an intended function as ``summarize'', there might be more than one reasonable mapping, that is, multiple functions (and thus outputs), given the same input. The sizes of these subspaces thus reflect the variety in how the model carries out the user's desired functions. 

The UIs illustrated in the figure show how they allow users to dive into the space at different points and also limit the reachable subspaces that can be explored by the user with that UI. As concrete examples, we illustrated conversational UIs, text input fields for free prompts, and toolbars. These approaches differ, for example, in the amount of involved context that is passed to the LLM, such as conversation history, the possibility to prompt instructions, and subsequent user interactions that involve the LLM's responses. These differences shape how these UIs support the user in accessing a desired function within the functional space. 
Based on the deconstructed UI examples in the figure, we analyze our UI implementations in light of functional space and functional flexibility:

Example A shows a conversational UI\revision{, similar to existing industry chat tools such as ChatGPT,} that offers a step-wise navigation of the functional space. Since the history of the conversation serves as context, with each utterance, the user iterates on the accessed function and thus pushes through the functional space until a satisfactory target function is reached. In our first prototype, multi-turn conversations were not implemented, yet some people asked if it was possible to reply to the AI, to iteratively push to other (refined) functions, for example, to adjust generations by making them longer (``do I have to write 'needs to be longer'?'') or summarizing a rather long suggestion (``summary of extended text'').

Example B shows a single prompt input in a text field\revision{, that is a zero-shot prompting interface as in Wordcraft \cite{yuan_wordcraft_2022}, and Script\&Shift \cite{siddiqui_scriptshift_2025}.} Similar to a conversational UI, these text fields offer users the opportunity to reach every point within the space of AI functions, but with only one try (assuming a basic version of this UI element without additional features). This free prompting offers a maximum of functional flexibility to the user and users have to carefully describe the intended function because this shapes the targeted subspace. Findings on the prompting behavior with our second prototype show that users specified parameters in their prompts to shape the target function (``Extend: write about \textit{(some meals)}, \textit{(that can be made with tomatoes)}'', and ``Reduce this text \textit{(by 50 words)}''; we put the parameters in parentheses/italic).

Example C shows a toolbar UI\revision{, similar to AI toolbars in industry solutions such as Grammarly.} This type of UI constrains the functional flexibility more than the examples before. Such toolbars offer pre-determined tools that capture commonly desired functions informed through user research. They thus provide a set of AI functions that the user can execute without entering a text prompt. Since these functions are pre-determined (e.g. via prompts hardcoded in the backend), they reach only certain functional subspaces, as defined by system builders. Based on the interactions with our second prototype, we found users to intuitively apply these AI functions, and chain them in various ways to explore the pre-determined subspaces (pre-determined AI functions accounted for 64.42\% of used functions, the others were used via free prompting). %

\subsection{Functional flexibility as a thinking tool}\label{sec:funcflex_thinking_tool}

We conclude with conceptual reflections and \revision{application} examples. Conceptually, our functional space is different from design spaces, which describe the options for (UI) designs, for example, structuring design requirements for co-creation \cite{lin_beyond_2023} or connecting HCI demands with model capabilities \cite{morris_design_2023}. In contrast, our conceptual lens describes how different UIs make AI functions accessible in different ways. This accounts for more than prompt-based interfaces \cite{subramonyam_bridging_2024}: As a key conceptual contribution, %
we offer a perspective for analyzing and comparing UIs that provide AI capabilities and that are as different as CUIs, traditional GUIs, and text-based prompting UIs.

\revision{It is possible to analyze further designs with this lens. For example, we discuss the interface from the application Script\&Shift by \citet{siddiqui_scriptshift_2025} through the lens of our construct by answering analytical, critical, and constructive questions from \cref{tab:generative-theory-questions}.}

\begin{figure*}[t]
    \centering 
    \includegraphics[width=0.75\textwidth]{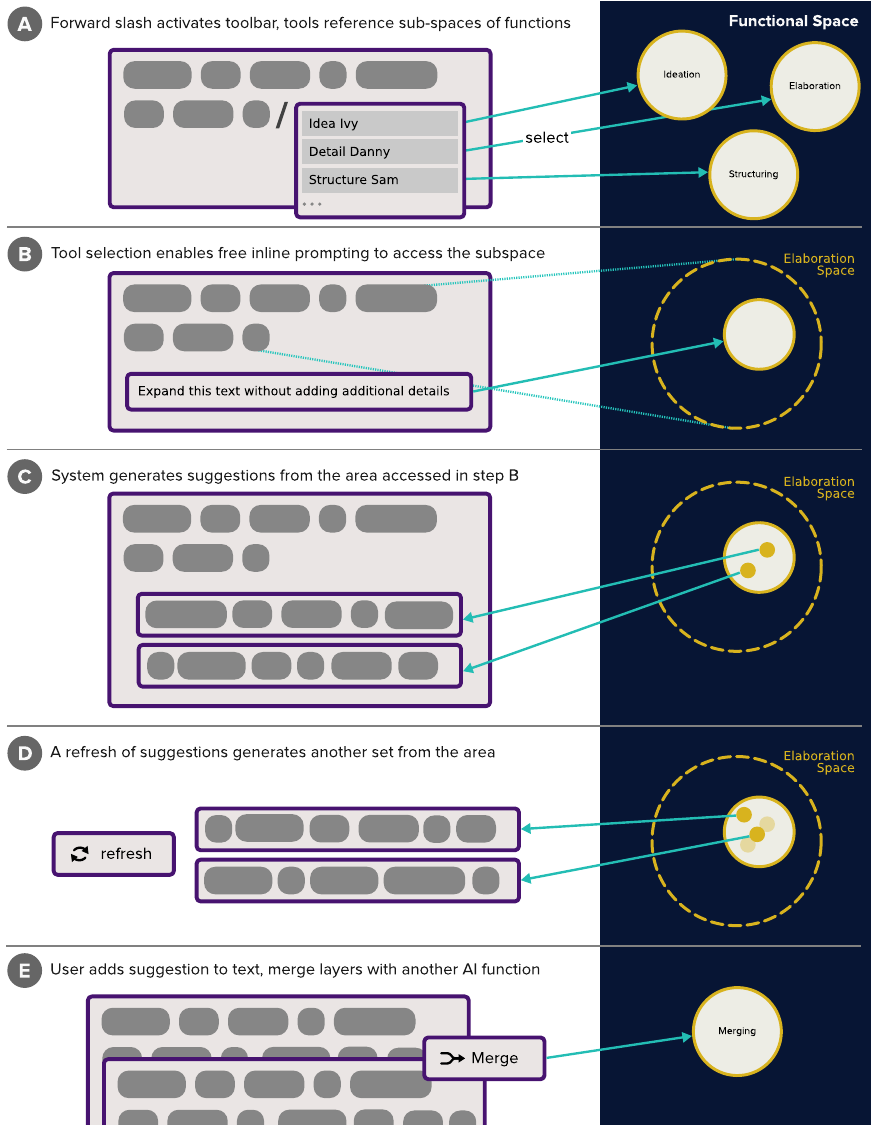}
    \caption{\revision{Illustration of how Script\&Shift \cite{siddiqui_scriptshift_2025} enabled accessing the functional space of the LLM. The illustration shows how (A) typing a forward slash triggers displaying an AI toolset. Users can select one of these functions. (B) Selecting a tool scopes a subspace, an inline free prompting text field allows users exploring this subspace. (C) The system generates a pair of suggestions. (D) Refreshing suggestions generates a new set. (E) Users can add the suggestion to the text. If multiple layers exist, they can be combined through a pre-determined merge function.}}
    \label{fig:funcflex-scriptshift}
\end{figure*}

\revision{In \cref{fig:funcflex-scriptshift}, we visualized how writing with the tool Script\&Shift enabled accessing the functional space of an LLM. We focus on their features ``text elaboration'' and ``merging layers''. (A) When people type a forward slash ``/'', an inline toolbar appears, offering six ``Writer's Friends'', such as Idea Ivy for ideation, and Detail Danny for elaborations. These Writer's Friends refer to subspaces of functions in the functional space. Users can select one of these Friends to scope a subspace. The tools confine the functional space of the LLM for the user. (B) An input field for free prompting appears below the paragraph, and gives the user maximum flexibility to navigate the subspace referenced through the previous selected Writer's Friend. The user is free to specify further instructions to the model, for example ``expand this text without adding addition details'' (cf. \cite{siddiqui_scriptshift_2025}). (C) By conditioning the LLM's generation on the user's entered prompt, the system generates two suggestions. The user can choose one of the two suggestions for including it into the original text. (D) A re-generation of the suggestions triggers the system to serve two other suggestions from the same subspace, as in the previous step. (E) With Scrip\&Shift, it is possible to write in multiple layers. Their concept of layers is similar to unconnected pages that can be interacted with on an infinite canvas space. Some Writer's Friends generate such pages as an output -- for example, Structure Sam generates two differently organized variants of the current page. Such pages can be merged into a single page by letting the LLM combine them.}

\revision{Dissecting the UI and the interaction flow in this way shows that the application involves a combination of an AI toolbar and an inline text input field for free prompting, similarly illustrated in example B in our \cref{fig:space}. Selecting a tool (Writer's Friend) from the AI toolset (list of Friends) scopes a subspace (low flexibility) and the free prompting UI element gives back maximum flexibility to navigate this subspace. Dragging and dropping one page onto another again scopes a pre-determined subspace (low flexibility) to let the LLM merge the two pages. This answers the analytical question from \cref{tab:generative-theory-questions} ``What UI elements are available with which degree of functional flexibility?''.}

\revision{The AI toolbar offers six pre-determined functions, which guide the user with decreased functional flexibility. However, some users might want to specify their own functions for maximum flexibility, i.e. free prompting without selecting one of the predefined Writer's Friends before. These users could benefit from defining their own Writer's Friend to reach their goals, and thus more freely explore the functional space. This answers the critical question ``Can users reach/identify their goals with the degree of functional flexibility offered by the system?''.}

\revision{Adding another trigger, such as typing a square ``\#'' could open an inline text input field for free-prompting own functions. These functions could be named and added to the toolbar, through checking a box next to the input field and adding a name. This way, users could add repeatedly used AI functions as further Writer's Friends. This free-prompting of functions (maximum flexibility) and making them available as reusable Writer's Friends (confined flexibility) would allow for dynamic switches of functional flexibility. This answers the constructive question ``Which UI elements could be added to a design to increase/decrease functional flexibility?''. }

\revision{Reflecting on Script\&Shift through the lens of Functional Flexibility enabled us to dissect the interaction analytically and to constructively critique the design. This demonstrates the value of Functional Flexibility as a thinking tool for designing human-AI interaction.}

More generally, we hope that our theoretical construct serves as a useful \revision{thinking tool} for researchers, designers, and engineers to analyze and reflect on how specific UI implementations support users in locating desired functions within the functional space provided by LLMs. Crucially, this provides a meaningful level of abstraction to comparatively discuss \textit{both} conversational and non-conversational UI choices. We deem this relevant and impactful in the current context, where many AI features in practice seem to predominantly consider conversational interaction.

\section{Discussion}\label{sec:discussion}

\subsection{Differentiating functional flexibility from other concepts in Human-AI interaction}

How does functional flexibility relate to other concepts relevant to Human-AI interaction, such as affordance, intent formulation, task decomposition, and mental models? We discuss this here:

\textit{Affordance} defines how UIs make their function apparent \cite{norman_design_2002}. Thus, UI elements should visibly signal the amount of functional flexibility they are offering to the user, for example the size of a text input field may indicate greater functional flexibility than a single line input field, or even a button. 

\textit{Formulating intent} in free text (prompting) is demanding for users \cite{zamfirescu-pereira_why_2023} but may offer the possibility to scope an intent with a higher degree of functional flexibility. However, decreased functional flexibility may allow users to express intentions more quickly if they are offered in the (limited set of the) UI as a tool. 

Such tools can also support \textit{decomposing a task} bottom up (e.g. first translating and then summarizing a text), as their presence in the UI already reveals these available AI functions as possible ``steps''. Higher functional flexibility instead allows for flexibly decomposing tasks (top-down) but imposes metacognitive demands on the user to do so \cite{tankelevitch_metacognitive_2024}.

Interacting with different degrees of functional flexibility may also influence the \textit{mental model}, that is, how the user thinks that the system works. A UI with high functional flexibility (e.g. prompt box, chatbot) may be more likely to be thought of as an agent, assistant, or partner -- with users leaving the mental model at that high level of abstraction. Conversely, UIs with lower functional flexibility (e.g. a toolbar) may result in more specific mental models because each individual UI element (e.g. button in the toolbar) can be more clearly linked to a more specific expected function and output. 

In summary, this short discussion in light of other concepts illustrates how functional flexibility can extend and refine our conceptual understanding of human-AI interaction. 

\subsection{Conversation is not all you need: ``regression to commands'' in repeated prompting}

We offered a conversational UI in study 1, based on the rich task delegations elicited in the survey. However, participants did not actually delegate tasks in this rich way. Instead, they went with command-like instructions and optimized their prompts down to a minimum, for example, writing ``extend'', or even more extreme, ``extd'', for delegating the AI to extend the marked text. 

We expect similar effects to show up in various prompting-based interaction contexts, not limited to writing. We propose to refer to this effect as \textit{``regression to commands''}. From the user perspective, this is optimizing or satisficing~\cite{simon1996artificial}, considering the trade-off of input effort vs result quality. Complementary, from a researcher perspective, it is something to be aware of in methodological considerations: We might elicit diverse prompts in one sample (e.g. our survey) because people can think of elaborate prompts. However, repeatedly eliciting prompts in an actual workflow reveals the (possibly much more narrow) ``point'' that is good enough to access that model functionality in interaction.

In contrast to the first study, in study 2, we offered direct tools for the repeated tasks, plus an optional free prompt input. In this setup, participants used the free prompt input selectively (35.58\% of AI triggers), but when they used it, they indeed entered more elaborate prompts, as reflected in Figure~\ref{fig:comparison_prompt_lenth}. 

Critically and constructively, this has strong implications for human-AI interaction research and design: Current AI is often integrated into products as a chatbot, heavily biased by ChatGPT's success, but this ignores a fundamental need in people's actual workflows. These often involve \textit{repeated} subtasks, which in turn require repeated access to specific functionality from the LLM's space of functions. As our conceptual lens highlights (\cref{fig:space}), CUIs are great to navigate the space towards a target function. However, CUIs are not ideal for repeatedly accessing the same target function, because this demands reentering prompts again and again. Instead, this repeated use is much better addressed by a toolbar and, more generally, by direct manipulation interfaces (e.g. \cite{Masson2024directgpt}) that allows users to access LLM functionality without text entry.

\subsection{The functional flexibility funnel: turning users' free prompts into future tools}

In another constructive application of functional flexibility, we propose to consider how the \textit{parallel} use of UI elements that provide different degrees of functional flexibility, such as free prompt inputs and toolbar-like UIs, can serve as a user research method to elicit functional user needs within a system even after its deployment \cite{amershi_guidelines_2019, weisz_design_2024}. 

With UI elements that provide specific AI functions, for example, the summary tool in our AI toolbar, it may happen that there are mismatches between the provided function and the user's expectation. This was the case, for example, when the summaries were too short or did not reflect important aspects of the original content. As a mitigation strategy, participants used the UI element for free prompt input to declare their own functions, such as ``summarize \textit{every section}'' and ``summarize text: \textit{4-5 sentences}''. In these examples, they parameterized the scope of the text document and the text length. In other prompts, they specified parameters for the tone and topic, such as ``rewrite to feel more \textit{natural}'', ``Write something about the \textit{taste of tomatoes}''. In doing so, they provided us with data about the actual function they would have liked to access.

Thus, a key insight here is that these parameters could be directly integrated into the UI. For example, a toolbar UI for extending or summarizing text could display a slider to adjust the text length. \textit{DynaVis}~\cite{vaithilingam_dynavis_2024} proposed dynamically generated widgets to realize a similar idea for prompting-based interaction with information visualizations. Beyond UI generation, the users' prompts could also be used to inform software updates. For example, system builders could use them to identify and inform \textit{fundamentally new} AI functions to be provided (as tools) in an update.

In this way, the application's UI could provide more and more refined tools for direct access to AI functions over its lifecycle. From an initially flexible space of functions accessed via prompt input, users guide the development towards a practically relevant toolset, accessed without prompting. In turn, (future) users are guided in their use and prompting needs by the already available tools. Metaphorically, we refer to this refinement of prompting needs as the \textit{functional flexibility funnel}. Note that this does not mean that flexibility is increasingly restricted -- the UI might always offer a free prompt option.

\subsection{The design of user interfaces shapes how users access functional flexibility}

Offering a maximum of functional flexibility to users might be tempting for designers and developers. For example, at first glance, a simple input field for free prompting seems to replace the need to design any other UI elements. We currently see this strategy in many products, such as ChatGPT or Microsoft Copilot -- their UIs are mainly built on and marketed as a promise of the flexibility of an open text box. However, applying our lens of functional flexibility critically, this \textit{shifts the responsibility for locating AI functions} completely to the user. 

Instead, designers and engineers could identify and implement low-flexibility UI elements as \textit{starting points} for users to access (some) AI functions more easily. Such pre-determined functions (e.g. offered as buttons) lower the barrier for using AI functions since users do not have to locate these functions in the space by themselves.

Indeed, study 2 revealed that participants were initially clueless about prompting and asked for examples (e.g. ``I wish for a list of examples, I was not sure if I was free about text input''). However, these participants referred to pre-determined AI functions for their first prompts: They input short commands that could possibly extend the toolbar, such as ``itemize'', or specifically extended existing functions, such as ``summarize 2-3 sentences less''.

Thus, pre-determined functions and their related UI elements can also serve as examples that set perceivable ``landmarks'' in the functional space: They reveal a subset of what is possible and thus \textit{shape user expectations about the whole space}.

Overall, our proposed perspective enables researchers, designers, and engineers to better assess and compare the various possibilities of UIs to shape functional flexibility. In particular, we expect that making functional flexibility an explicit consideration in the design of interactive AI systems helps reveal design opportunities beyond (or in combination with) conversational UIs and simple prompt boxes. %

\subsection{Limitations \revtwo{and reflections on a multi-year research project}}

We followed a mixed-methods approach and observed participants' screens via video calls. This approach offered insights we otherwise would have missed. %
For example, observations revealed how participants chained AI functions. People also shared their thoughts \textit{during} writing, which not only provides a glimpse at their thinking processes but also helps us to interpret the findings from interaction logging (cf. \cite{bhat_interacting_2023}). %
Overall, this investment comes with typical tradeoffs of qualitative research, such as favoring detailed observation over a larger sample and risking that observations might influence people's behavior.

We do not claim to have achieved an ``optimal'' implementation of the AI features and this would be a moving target anyway, since model development advances quickly. ``Perfect'' text generation was not required for our research here, as long as AI-generated text could be realistically included in writing workflows. This bar was met for the tasks in our studies, based on our own assessments and people's perception. %

\revtwo{Finally, this article reports on empirical work spanning several years, leading to the presented theoretical construct (\cref{fig:framing}). Here, we reflect on the dynamic developments of generative AI in interactive applications during this timeframe:}

\revtwoflo{Our prototypes reflect the model capabilities available at the time of their development, and it is worth reflecting on how model progress since then affects implementation choices and findings.
Our first prototype relied on a separate intent-recognition component (based on RASA) to route user comments to task-specific models for summarizing, translating, and extending text. The interactive comments did not support multi-turn conversation. Current LLM APIs offer conversational interaction by default and can reason about which function to invoke without a dedicated classifier, which reduces the implementation effort such a system would require today. While conversational ability was already present in GPT-3, native reasoning about function invocation became broadly available only with later models (e.g., function-calling-capable GPT-3.5). Model capabilities have generally improved to the point that a single general-purpose LLM could now replace the multiple smaller, task-specific models (T5, Opus-MT, GPT-Neo/GPT-3) we combined in our prototypes.}

\revtwoflo{Beyond technical capability, user experience with AI has also shifted: With the public availability of ChatGPT in late 2022, people may have gained more experience in delegating tasks to AI than the participants of our 2021 study, which may influence how users express their intent today. However, our second study was conducted shortly after the time ChatGPT became publicly available, and our core findings do not depend on model quality or user familiarity with chatbots. The ``regression to commands'' we observed in Study 1 stems from the effort of repeatedly expressing the same intent, not from model limitations, and we expect this would also happen today, regardless of how capable or familiar the underlying model is. Indeed, recent work on direct manipulation interaction for LLMs and generative UI is motivated by this (e.g.~\cite{Masson2024directgpt, vaithilingam_dynavis_2024}). Similarly, the value of pre-specified AI functions (toolbars) alongside custom prompting is rooted in the discoverability of, and cognitive load of locating, a desired function within the almost unlimited functional space an LLM offers. As a challenge, this persists \cite{tankelevitch_metacognitive_2024, zamfirescu-pereira_why_2023}, and arguably grows, as models become more capable and support more functions. Our theoretical construct of functional flexibility synthesizes these insights from both studies together with related work: The UI defines how users can express their intents and to what degree they can access the LLM's functional space. This thinking tool makes it possible, as demonstrated, to dissect how interaction with LLMs is realized across very different UI paradigms, independent of any particular model's capability at a given point in time.}

\section{Conclusion}

We investigated interaction with text models in multiple studies over four years. Within this period falls the moment when writing with conversational AI became widely popular. Our work includes a survey on text editor usage and task delegation to AI without the ``legacy bias'' of ChatGPT and two user studies: one with a conversational UI and another with a toolbar UI, offering pre-determined AI functions alongside custom prompt input. 
When interacting with the conversational UI, people used short command-like instructions for task delegation, in contrast to the more elaborate prompts elicited in our survey. With the toolbar UI providing \textit{both} pre-specified AI tools and flexible prompting, people used and combined the tools for specific tasks, while they invested into free prompting to realize custom functions.

Synthesizing \revthreeflo{observations from these studies} with contributions from related work \revthreeflo{for theory-building}, we proposed the new theoretical construct and perspective of \textit{functional space} and \textit{functional flexibility}. We reflected on our prototypes and related work in light of functional flexibility, considering analytical, critical, and constructive perspectives. \revthreeflo{Our theoretical construct} reveals and makes explicit a new key mechanism by which ``interaction with LLMs'' can be construed in HCI: accessing a desired target function through suitable UIs. In this light, UIs serve the key role of granting users access to the functional space of LLMs and their design shapes how users can do so, leading to different degrees of functional flexibility. With our work, we thus contribute to function-oriented design of human-AI interaction with generative AI. 

Although it is tempting to offer a maximum of flexibility to users by letting them decide on relevant functionality via prompting themselves, our investigation \revthreeflo{suggests} that effectively integrating AI into UIs requires actively thinking about, and designing for, providing features with the right amount(s) of functional flexibility.

\bibliographystyle{ACM-Reference-Format}
\bibliography{bibliography}

\pagebreak

\section{Appendix}
\label{appendix}

\subsection{Questionnaires}
\label{apx:questionnaires}
The complete questionnaire of our survey can be found in our OSF repository\footnote{\url{https://osf.io/qh46n}} alongside the other questionnaires in our studies. Since the latter were rather short questionnaires, we also put them into this appendix. %

\subsection{Task Briefing}
\label{apx:task-briefing}

\subsubsection{Opener study 1:}

Here, you will use an online text editor with AI skills.

The AI skills support you with summarizing, extending, and translating text within the editor. By commenting parts of the text with to-dos an AI author will take over these tasks. See the screenshots below for a short demonstration.

\begin{figure}[ht]
    \centering
    \subfloat[\centering Step one]{{\includegraphics[width=0.4\textwidth]{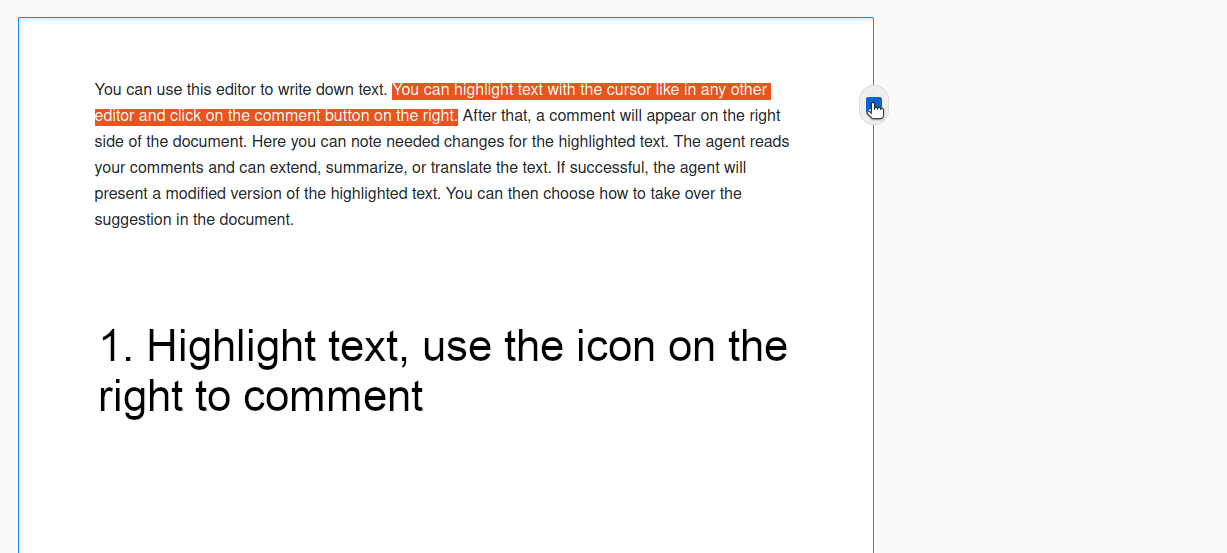} }}
    \qquad
    \subfloat[\centering Step two]{{\includegraphics[width=0.4\textwidth]{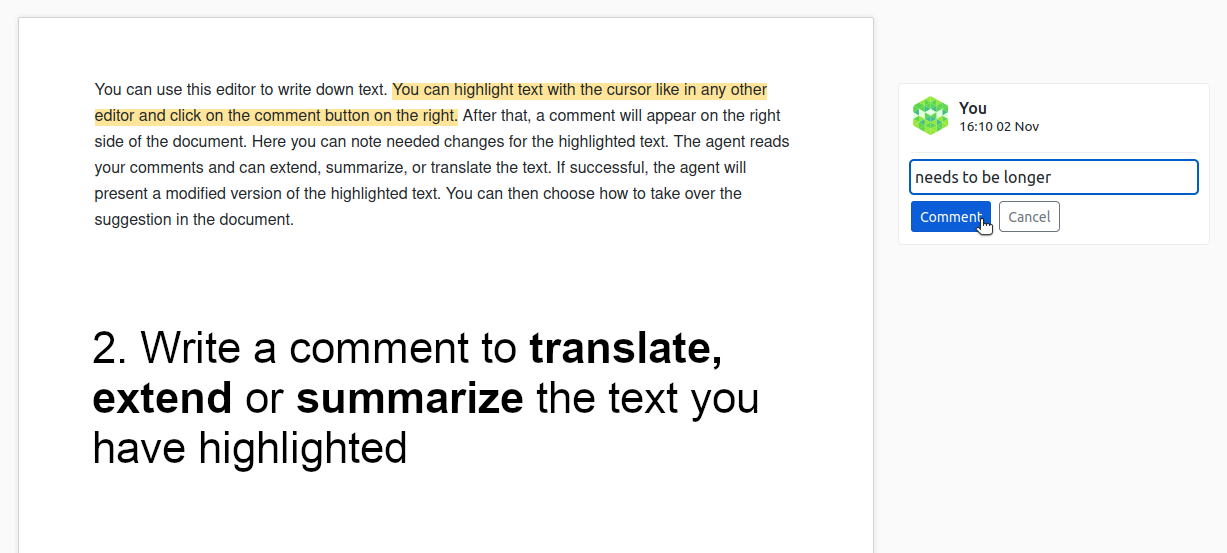} }}
    \qquad
    \subfloat[\centering Step three]{{\includegraphics[width=0.4\linewidth]{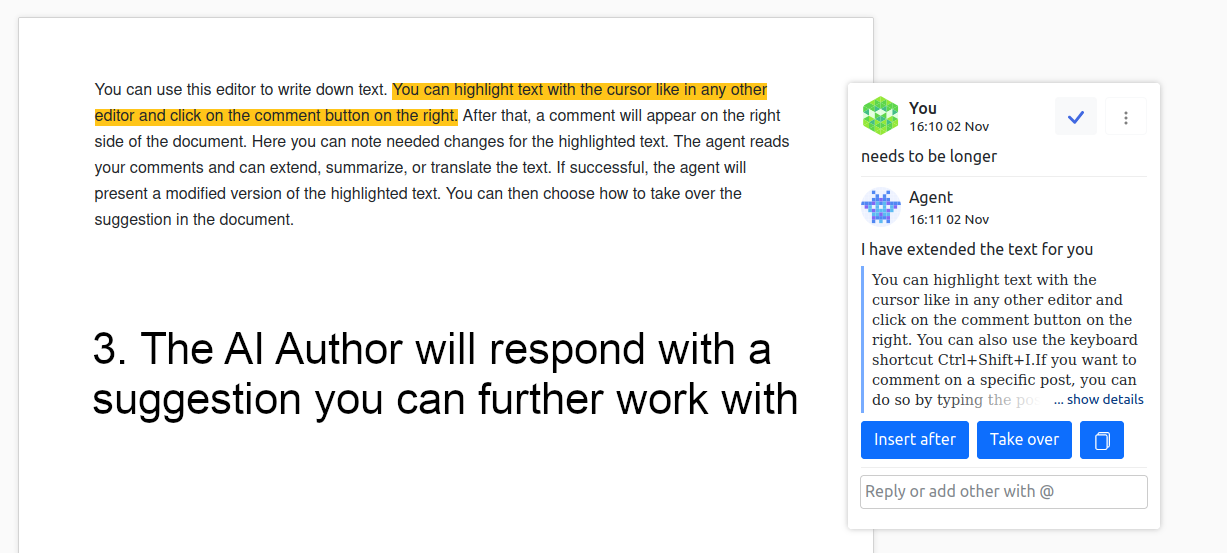} }}
    \caption{Stills from the video we presented to participants within the task briefing of study 1. The stills explain in three steps how a writing task can be delegated to AI.}
    \Description{This figure shows three screenshots taken from a video we presented to participants as part of the task briefing. The stills explain in three steps how the editors can be used to delegate a writing task to AI through interactive comments.}
    \label{fig:briefing_1_stills}
\end{figure}

\textit{Video:} Stills of the video can be seen in \Cref{fig:briefing_1_stills}.

\subsubsection{Opener study 2:}

Here, you will use an online text editor with an AI toolset.

The AI toolset supports you with summarizing, extending, and translating text within the editor. Furthermore, you can specify an AI action on your own using the prompt tool. By selecting parts of the text and using the AI tool from the floating toolbar, an AI action will be applied to your text. See the video below for a short demonstration.

\begin{figure}[ht]
    \centering
    \subfloat[]{{\includegraphics[width=0.38\textwidth]{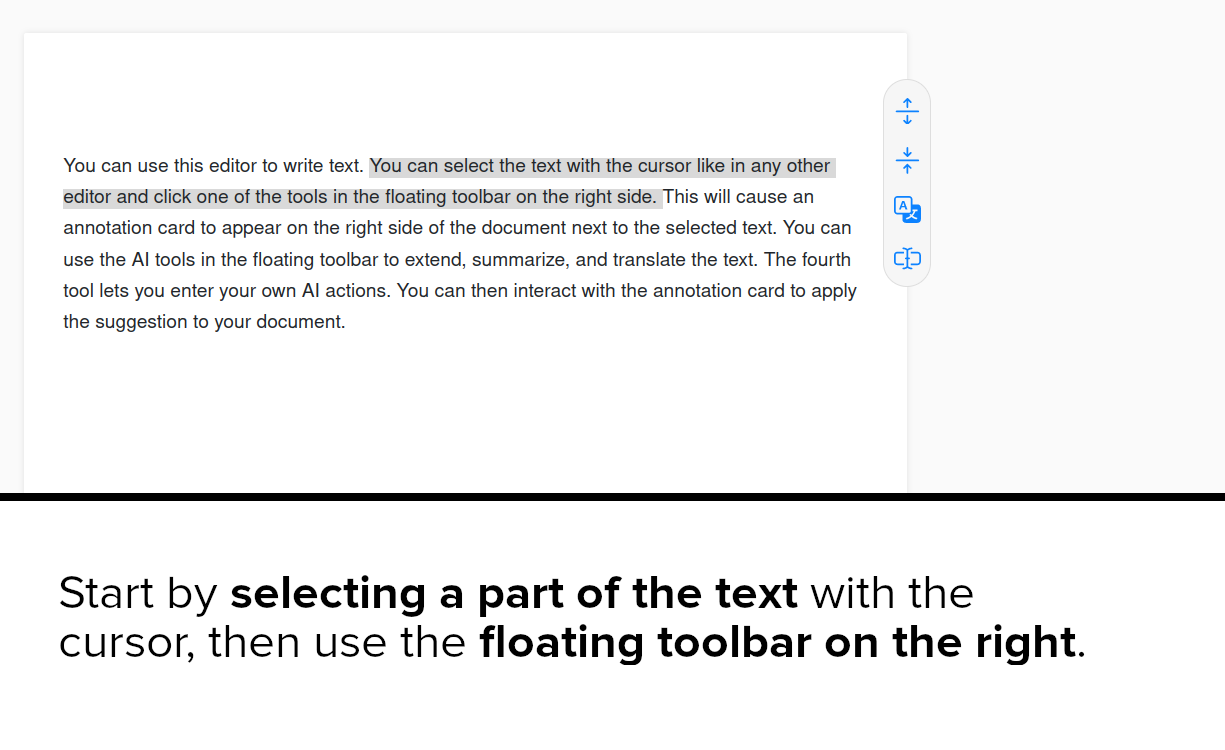} }}
    \qquad
    \subfloat[]{{\includegraphics[width=0.38\textwidth]{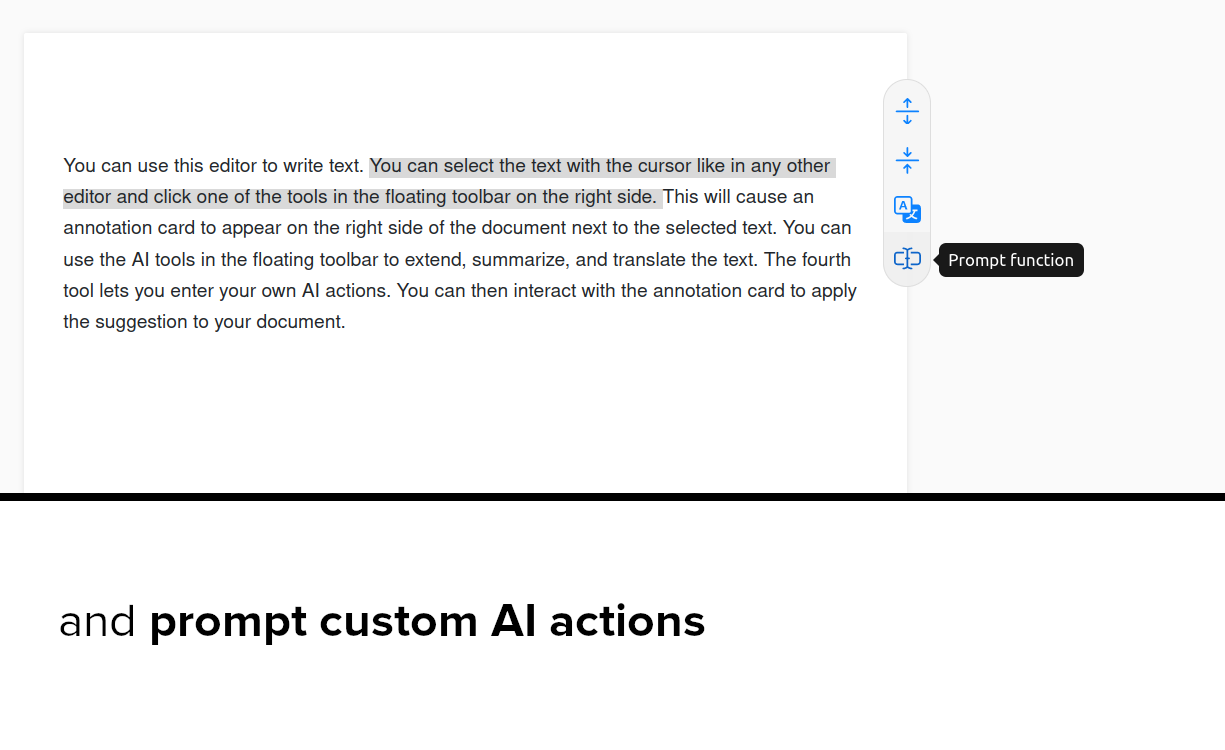} }}
    \qquad
    \subfloat[]{{\includegraphics[width=0.38\textwidth]{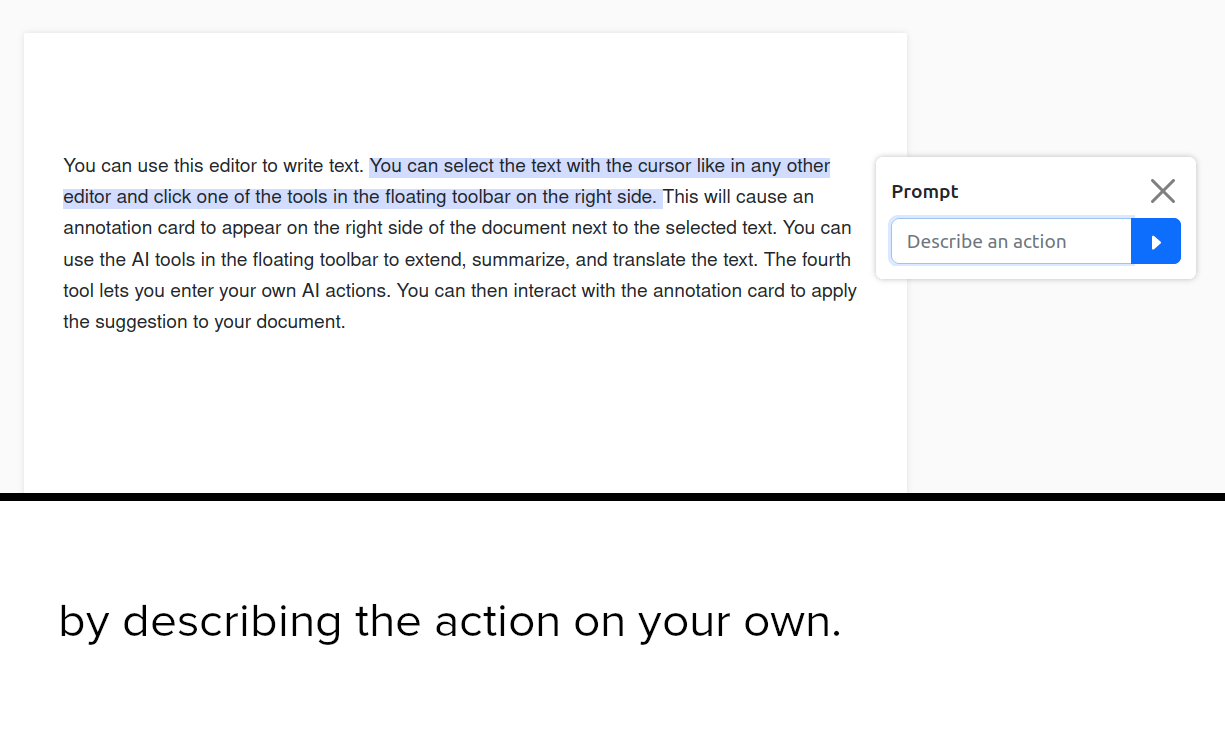} }}
    \qquad
    \subfloat[]{{\includegraphics[width=0.38\textwidth]{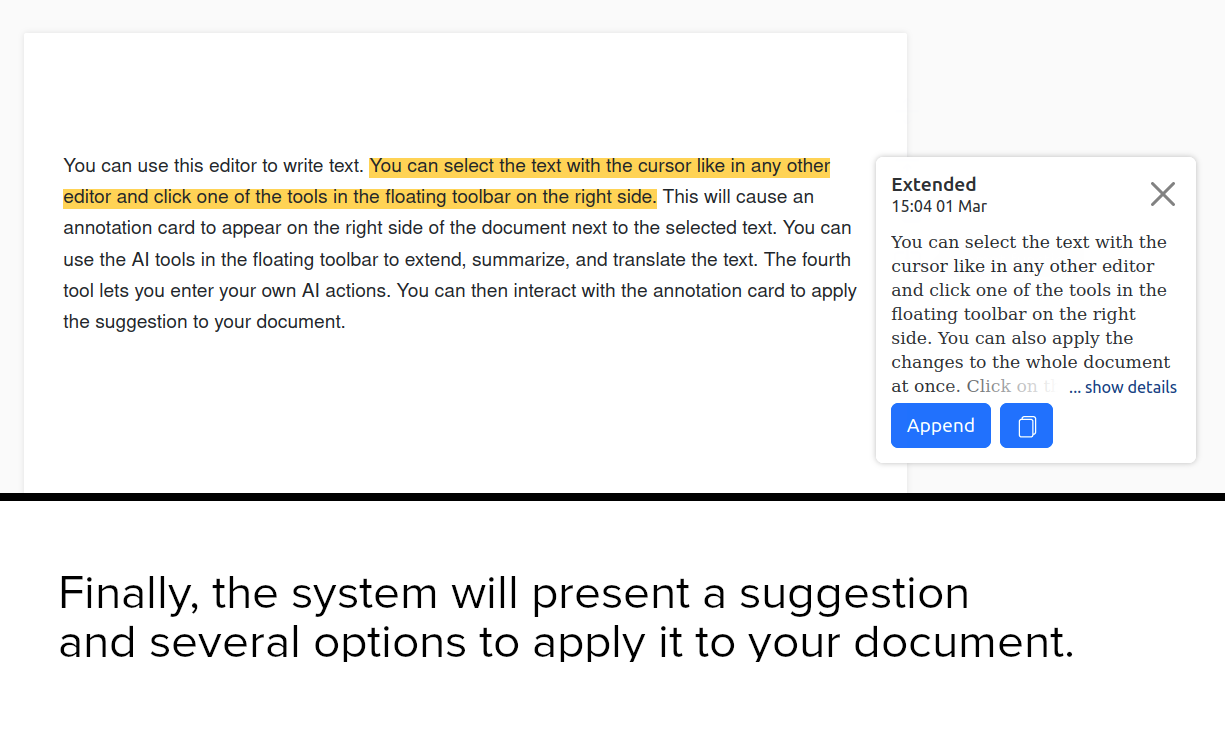} }}
    \caption{Stills from the task briefing video we presented to participants in study two. The stills show in four steps how an AI function can be used from the toolbar UI (pre-determined functions, and free prompting of functions).}
    \Description{This figure shows screenshots taken from a descriptive video we presented to participants of our second study. The stills present how pre-determined AI functions and how users can prompt their own functions.}
    \label{fig:briefing_2_stills}
\end{figure}

\begin{figure*}[]
    \centering
    \subfloat[\centering The responses on the SUS questionnaire in our first user study showed a a mean score of 83.5, that is comparable to an ``excellent usability'' \cite{bangor_empirical_2008}]{{\includegraphics[width=1\linewidth]{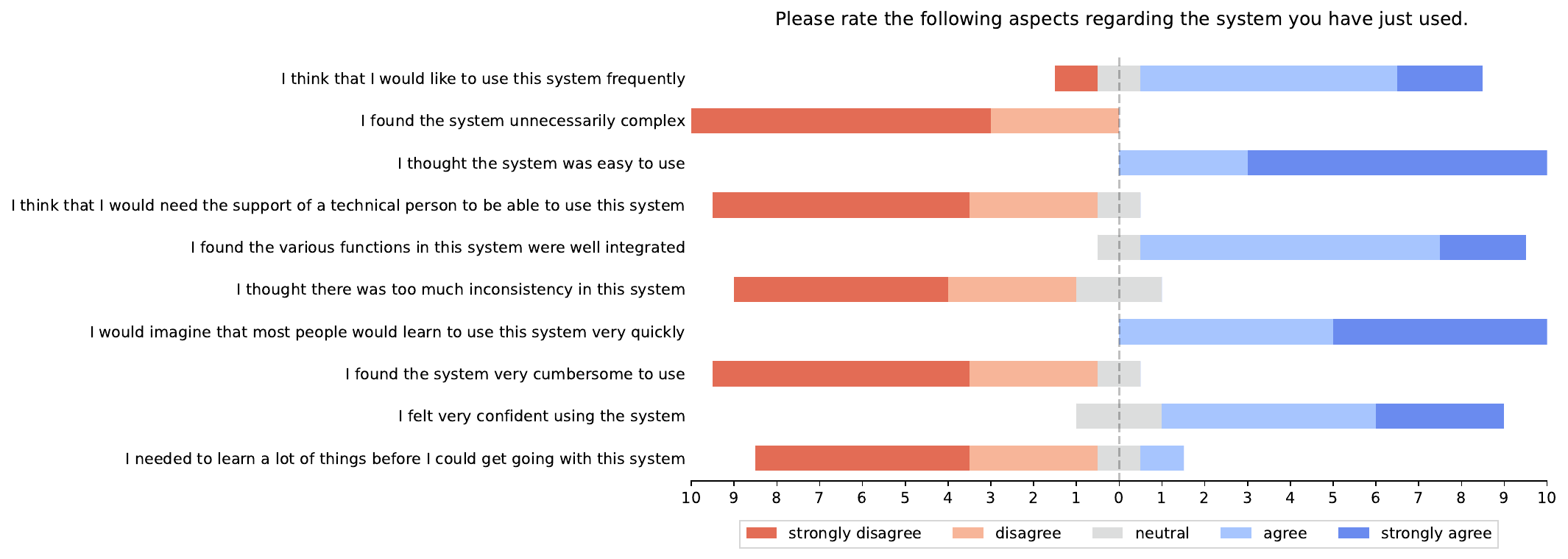} } \label{fig:cui_sus_ratings} }
    \qquad
    \subfloat[\centering Ratings on questions, asking for system usage and topic familiarity.]{{\includegraphics[width=1\linewidth]{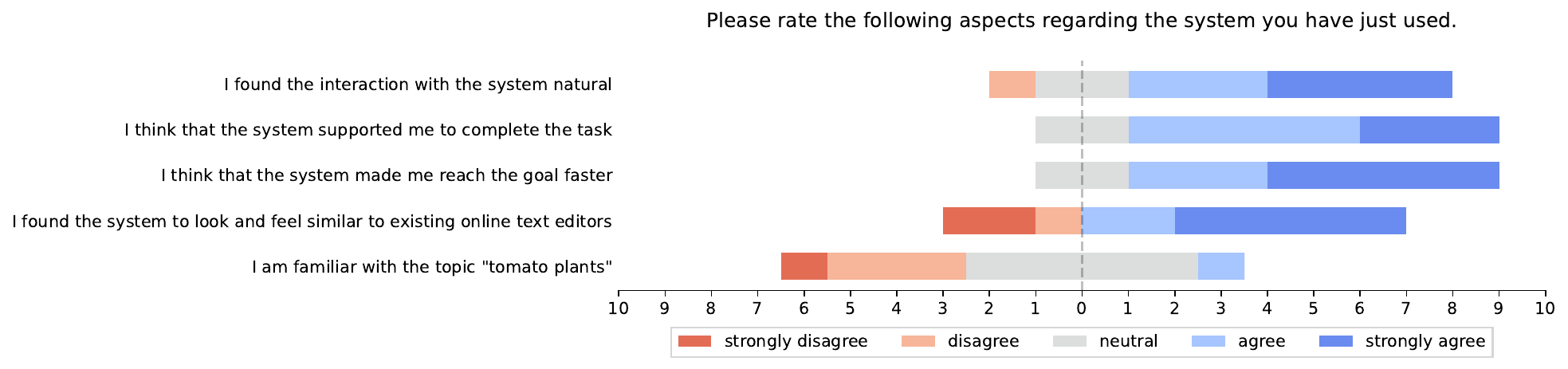} } \label{fig:cui_additional_ratings} }
    \caption{Divergent bar charts of responses on the SUS and additional ratings in study 1.}
    \Description{This figure shows two divergent bar charts. The first chart displays the responses on the SUS questionnaire in study 1. The second chart displays the responses on additional questions such as naturalness and efficiency.}
    \label{fig:cui_ratings_overview}
\end{figure*}

\begin{figure*}[]
    \centering
    \subfloat[\centering Overview of responses on the SUS questionnaire of our second user study, with a mean score of 87.29, that is comparable to an ``excellent usability'' \cite{bangor_empirical_2008}. The score is slightly higher than in study 1.]{{\includegraphics[width=1\linewidth]{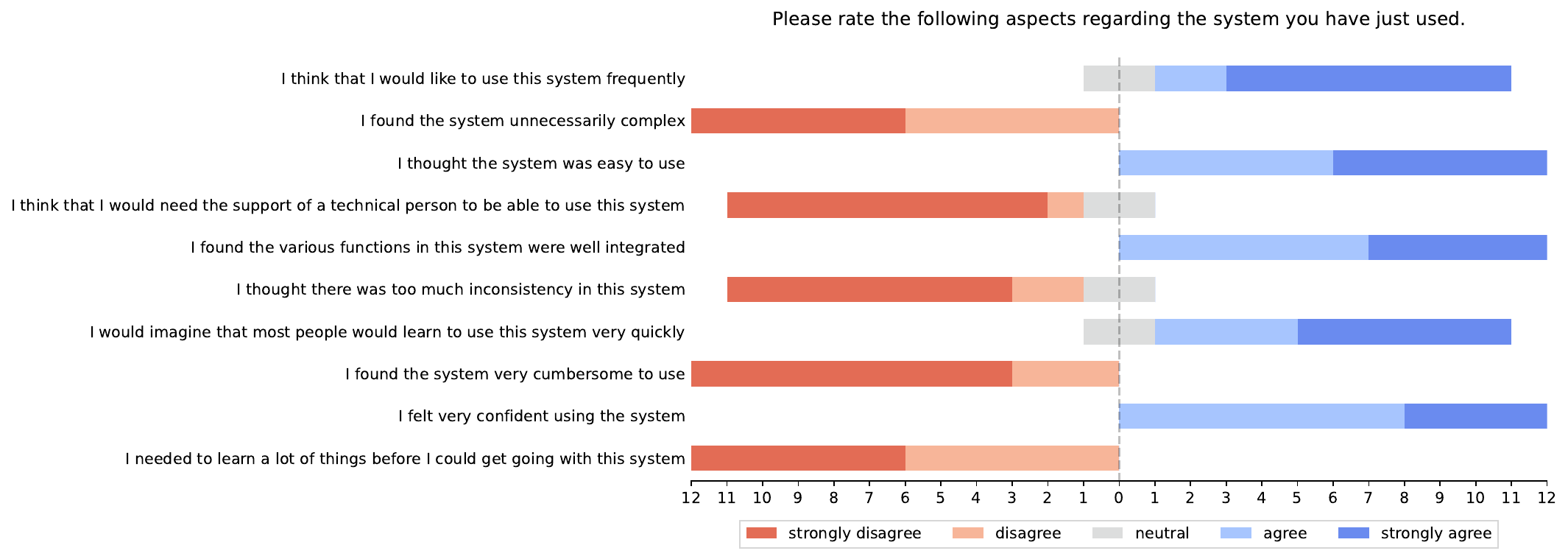} } \label{fig:tools_sus_ratings} }
    \qquad
    \subfloat[\centering Ratings on user perception in the seconds user study, asking for tool usage, perceived control, authorship, and topic familiarity]{{\includegraphics[width=1\linewidth]{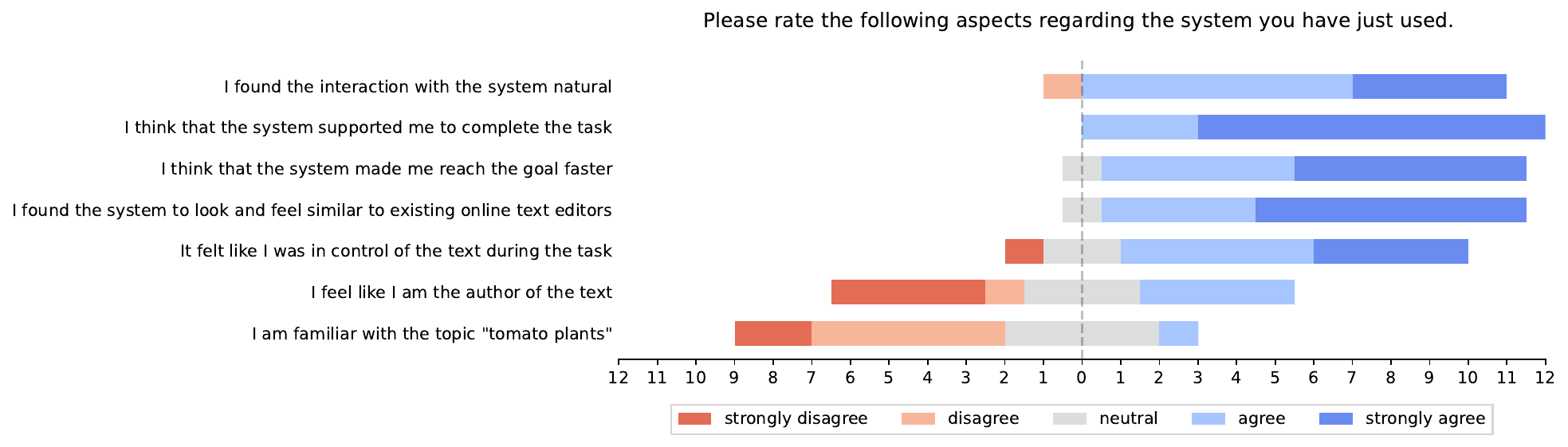} } \label{fig:tools_additional_ratings} }
    \caption{Divergent bar charts of responses on the SUS and additional ratings in study 2.}
    \Description{This figure shows two divergent bar charts. The first chart displays the responses on the SUS questionnaire in study 2. The second chart displays the responses on additional questions such as naturalness and efficiency, but also authorship and control.}
    \label{fig:tools_ratings_overview}
\end{figure*}

\begin{figure*}[]
    \centering
    \includegraphics[width=0.5\textwidth]{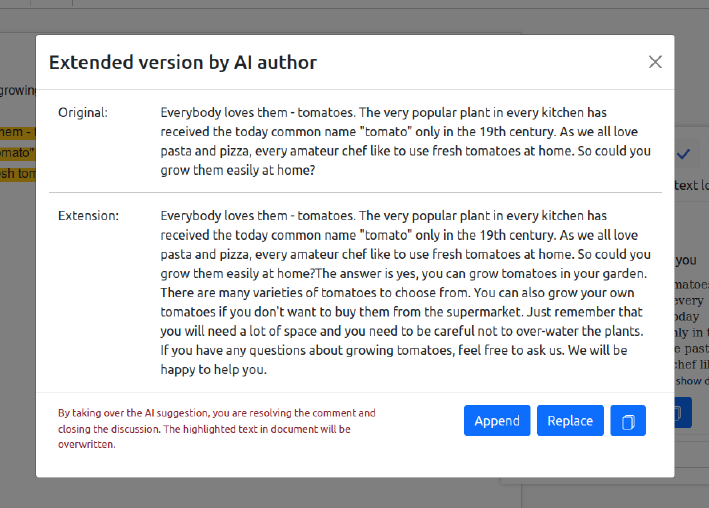}
    \caption{Screenshot of the detail view (``diff view'') that enabled users to compare their original selected text and the AI's response/suggestion.}
    \label{fig:prototype_1_diffview}
    \Description{This figure shows a screenshot taken from our prototype of study 1. It shows a modal UI that contains a diff view. Participants could use this diff view to compare marked text with AI suggestions and were presented options to accept the suggestion. The diff view is vertical, and buttons to accept the suggestions are placed at the bottom of the modal.}
\end{figure*}

\begin{figure*}[]
    \centering
    \includegraphics[width=0.5\textwidth]{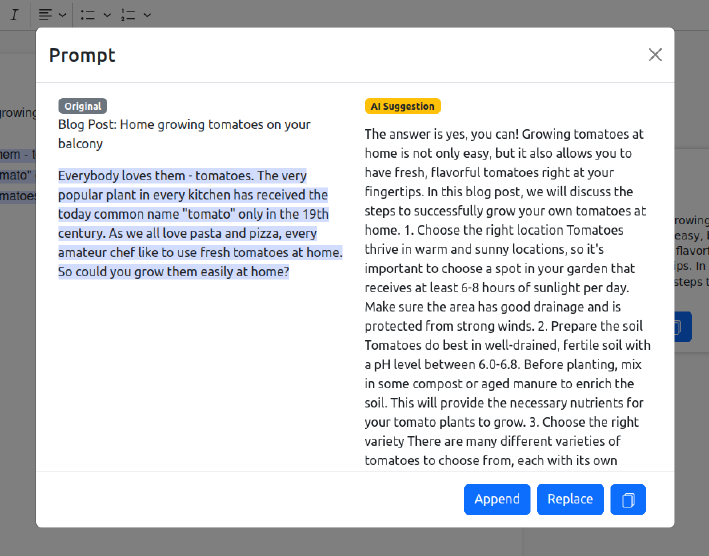}
    \caption{Screenshot of the improved detail view for AI suggestions/responses in our second prototype. The original text and AI suggestion were presented in a layout with two rows, which we then switched to two columns, as shown here.}
    \label{fig:prototype_2_diffview}
     \Description{This figure shows a screenshot taken from our prototype of study 2. It shows a modal UI that contains a diff view. Participants could use this diff view to compare marked text with AI suggestions, and were presented options to accept the suggestion. The diff view is horizontal, and buttons to accept the suggestions are placed at the bottom of the modal.}
\end{figure*}

\textit{Video:} Stills of the video can be seen in \Cref{fig:briefing_2_stills}.

\subsubsection{Remaining task briefing:}

You will have to write an informal blog post for a gardening blog. The main topic is tomato plants. In the same document, you will have to write a short summary about your blog post for impatient readers, known as "wrap up" or "too long did not read".

\textbf{Information on the task:}

\textbf{Task:} Write a blog post and a summary of that for a gardening blog.

\textbf{Topic:} Tomato plants.

\textbf{Language:} English.

\textbf{Text length:} The blog post must have a length between 300 - 350 words. Additionally, your summary should be concise with a length of three to five sentences maximum.

\textbf{Quality:} The quality of the text should be moderate/reasonable. It must not be perfect, yet you should avoid making a lot of mistakes. Try to be efficient and effective at the same time! The task does not assess your writing skills.

\textbf{Time:} You should try to finish the task within 20 - 30 minutes.

\textbf{Keywords:} The blog post should include the keywords as follows: Tomato, origin, summer, water, balcony, growing, fruit, taste.

\textbf{Resources:} Your are allowed to use the following resources for writing your blog post:

\begin{itemize}
    \item https://de.wikipedia.org/wiki/Tomate
    \item https://www.mein-schoener-garten.de/pflanzen/gemuse/tomaten
    \item https://www.gartentipps.com/tomaten-auf-dem-balkon-ziehen-wertvolle-tipps-zum-anbau.html
    \item https://www.plantura.garden/gartentipps/gemuseratgeber/tomaten-anbau-auf-terrasse-und-balkon
    \item https://www.poetschke.de/beratung/tomate-ratgeber/
    \item https://www.native-plants.de/blog/tomatenpflanzen-selbst-ziehen/
\end{itemize}

\clearpage

\subsection{Supplementary figures and tables}

In the remaining appendix, we append all supplementary figures and tables.

\begin{table*}[t]
\caption{Questions asked as part of study 1: A) system usability scale, B) system ratings, C) feedback, D) demographics.}
\Description{This table lists the questions we asked in study 1. The table contains the questions and the response types.}
\label{tab:study1-questionnaire}
\small
\begin{tblr}{X[1,l]X[12,l]X[6,l]}
\toprule
\textbf{ID} & \SetCell[r=1]{l}{\textbf{Question}} & \SetCell[r=1]{l}{\textbf{Response type}} \\ \hline
A1     & I think that I would like to use this system frequently                        & \SetCell[r=10]{m} 5-point Likert Rating         \\
A2     & I found the system unnecessarily complex                        &       \\
A3     & I thought the system was easy to use                        &           \\
A4     & I think that I would need the support of a technical person to be
able to use this system                       &           \\
A5     & I found the various functions in this system were well integrated                        &           \\
A6     & I thought there was too much inconsistency in this system                        &          \\
A7     & I would imagine that most people would learn to use this system
very quickly                        &           \\
A8     & I found the system very cumbersome to use                        &           \\
A9     & I felt very confident using the system                        &          \\
A10     & I needed to learn a lot of things before I could get going with this
system                       &           \\

\hline

B1     & I found the interaction with the system natural                        &  \SetCell[r=4]{m} 5-point Likert Rating        \\
B2     & I think that the system supported me to complete the task                        &        \\
B3     & I think that the system made me reach the goal faster                       &          \\
B4     & I found the system to look and feel similar to existing online text
editors                       &          \\

\hline

C1     & I think that I would like to use this system frequently                        & Free text         \\
C2    & What was negative about the system?                        & Free text          \\
C3     & I am familiar with the topic "tomato plants"                        & 5-point Likert Rating          \\

\hline

D1     & Please enter your age                       & Number         \\
D2     & What gender do you identify as?                       & Selection
(Woman, Man, Non-binary,
Prefer not to say,
Prefer to self-describe)        \\
D3     & What is your profession / occupation / job title?                        & Free text         \\
D4     & How well do you speak English?                        & \SetCell[r=2]{m} Selection
(No knowledge of \textit{[English / German]}, Speak poorly (beginner knowledge), Fairly well (intermediate knowledge), Well (advanced knowledge), Very well (proficient in English), Native speaker)        \\
D5     & How well do you speak German?                        &        \\
\bottomrule
\end{tblr}
\end{table*}

\begin{table*}[ht]
\caption{Questions asked as part of study 1: A) previous experience, B) system ratings, C) system usability scale, D) agency, E) feedback, F) demographics}
\Description{This table lists the questions we asked in study 2. The table contains the questions and the response types.}
\label{tab:study2-questionnaire}
\small
\begin{tblr}{X[1,l]X[12,l]X[6,l]}
\toprule
\textbf{ID} & \SetCell[r=1]{l}{\textbf{Question}} & \SetCell[r=1]{l}{\textbf{Response type}} \\ \hline

A1     & Please list all previous experiences you have in writing with generated
text.                        & Multiple selection (none, Writing with word or sentence suggestions, Writing with auto-completion, Writing with auto-correction, Using the smart reply feature, Using ChatGPT, other) \\

\hline

B1     & I found the interaction with the system natural                        &  \SetCell[r=4]{m} 5-point Likert Rating        \\
B2     & I think that the system supported me to complete the task                        &        \\
B3     & I think that the system made me reach the goal faster                       &          \\
B4     & I found the system to look and feel similar to existing online text editors                       &          \\

\hline

C1     & I think that I would like to use this system frequently                        & \SetCell[r=10]{m} 5-point Likert Rating         \\
C2     & I found the system unnecessarily complex                        &       \\
C3     & I thought the system was easy to use                        &           \\
C4     & I think that I would need the support of a technical person to be
able to use this system                       &           \\
C5     & I found the various functions in this system were well integrated                        &           \\
C6     & I thought there was too much inconsistency in this system                        &          \\
C7     & I would imagine that most people would learn to use this system
very quickly                        &           \\
C8     & I found the system very cumbersome to use                        &           \\
C9     & I felt very confident using the system                        &          \\
C10     & I needed to learn a lot of things before I could get going with this
system                       &           \\

\hline

D1     & It felt like I was in control of the text during the task                       &  \SetCell[r=2]{m} 5-point Likert Rating        \\
D2     & I feel like I am the author of the text                        &        \\

\hline

E1     & I am familiar with the topic "tomato plants"                        & 5-point Likert Rating          \\
E2     & What was negative about the system?                        & Free text         \\
E3    & What was positive about the system?                       & Free text          \\

\hline

F1     & Please enter your age                       & Number         \\
F2     & What gender do you identify as?                       & Selection
(Woman, Man, Non-binary,
Prefer not to say,
Prefer to self-describe)        \\
F3     & What is your profession / occupation / job title?                        & Free text         \\
F4     & How well do you speak English?                        & \SetCell[r=2]{m} Selection
(No knowledge of \textit{[English / German]}, Speak poorly (beginner knowledge), Fairly well (intermediate knowledge), Well (advanced knowledge), Very well (proficient in English), Native speaker)        \\
F5     & How well do you speak German?                        &        \\
\bottomrule
\end{tblr}
\end{table*}

\begin{table*}[]
\caption{Results of coded ``Goals'' participants had for delegating a task to AI in a writing stage. We discovered a desire for inspiration in the outlining stage, whereas participants aimed for efficiency and improving a text in the drafting and revising stages.}
\Description{This table shows results of coded ``Goals'' participants had for delegating a task to AI in a writing stage. The table lists the stages and counts of goals, such as inspiration and efficiency.}
\label{tab:writing-stages-goals}
\small
\begin{tblr}{lrrrrr}
\toprule
\textbf{Writing Stage} & \SetCell[r=1]{l}{\textbf{Inspiration}} & \SetCell[r=1]{l}{\textbf{Efficiency}} & \SetCell[r=1]{l}{\textbf{Learning}} & \SetCell[r=1]{l}{\textbf{Improve Text}} & \SetCell[r=1]{l}{\textbf{Coordination}} \\ \hline
Outlining     & 102                             & 159                            & 113                          & 49                               & 15                               \\
Drafting      & 51                              & 207                            & 73                           & 161                              & 8                                \\
Revising      & 12                              & 210                            & 33                           & 179                              & 4                                \\
Coordinating  & 8                               & 53                             & 28                           & 77                               & 204                            \\
\bottomrule
\end{tblr}
\end{table*}

\begin{table*}[]
\caption{Results of coded ``Styles'' participants applied for delegating a task to AI in a writing stage. Most often occurred an imperative style, followed by asking questions.}
\Description{This table shows results of coded ``Styles'' participants applied for delegating a task to AI in a writing stage. The table lists the stages and counts of appearing styles, such as imperative and question.}
\label{tab:writing-stages-styles}
\small
\begin{tblr}{lrrrr}
\toprule
\textbf{Writing Stage} & \SetCell[r=1]{l}{\textbf{Imperative}} & \SetCell[r=1]{l}{\textbf{Question}} & \SetCell[r=1]{l}{\textbf{Direct Input}} & \SetCell[r=1]{l}{\textbf{Statement}} \\ \hline
Outlining     & 196                            & 94                           & 1                                & 88                            \\
Drafting      & 245                            & 80                           & 0                                & 34                            \\
Revising      & 250                            & 87                           & 0                                & 24                            \\
Coordinating  & 252                            & 83                           & 3                                & 14                           \\
\bottomrule
\end{tblr}
\end{table*}

\begin{table*}[]
\caption{Results of coded ``Roles'' participants assigned to AI for delegating a task in a writing stage. We most often discovered participants to refer to AI as a co-author in the outlining stage. With advancing stages, they less often refer to AI as a co-author, but more often as an assistant.}
\Description{This table shows results of coded ``Roles'' participants assigned to AI for delegating a task in a writing stage. The table lists the stages, and counts of coded roles, such as co-author and advisor.}
\label{tab:writing-stages-roles}
\small
\begin{tblr}{lrrrr}
\toprule
\textbf{Writing Stage} & \SetCell[r=1]{l}{\textbf{Co-Author}} & \SetCell[r=1]{l}{\textbf{Advisor}} & \SetCell[r=1]{l}{\textbf{Assistant}} & \SetCell[r=1]{l}{\textbf{Editor}} \\ \hline
Outlining     & 143                           & 66                          & 150                           & 22                         \\
Drafting      & 98                            & 20                          & 195                           & 47                         \\
Revising      & 71                            & 25                          & 231                           & 39                         \\
Coordinating  & 24                            & 7                           & 279                           & 43     \\
\bottomrule
\end{tblr}
\end{table*}

\begin{table*}[]
\caption{Results of coded ``Styles'' participants applied for delegating different tasks to AI. We focus on the three AI capabilities: summarize, extend, and translate. We discovered a majority of elicited inputs to have an imperative style.}
\Description{This table shows results of coded ``Styles'' participants applied for delegating a certain AI capability. The table lists the AI capabilities and counts of coded styles, such as imperative and question.}
\label{tab:ai-skills-general-style}
\small
\begin{tblr}{lrrrr}
\toprule
\textbf{AI Capability}              & \SetCell[r=1]{l}{\textbf{Imperative}} & \SetCell[r=1]{l}{\textbf{Question}} & \SetCell[r=1]{l}{\textbf{Direct Input}} & \SetCell[r=1]{l}{\textbf{Statement}} \\ \hline
Revise                 & 158                            & 18                           & 0                                & 16                            \\
\textbf{Summarize}     & \textbf{143}                   & \textbf{10}                  & \textbf{2}                       & \textbf{2}                    \\
Format                 & 195                            & 10                           & 0                                & 4                             \\
Shorten                & 152                            & 17                           & 1                                & 4                             \\
\textbf{Extend}        & \textbf{135}                   & \textbf{19}                  & \textbf{0}                       & \textbf{13}                   \\
Highlight              & 166                            & 13                           & 4                                & 3                             \\
Cut                    & 178                            & 8                            & 1                                & 3                             \\
Complete               & 139                            & 8                            & 13                               & 1                             \\
Draft                  & 137                            & 9                            & 10                               & 17                            \\
Inspire                & 155                            & 12                           & 3                                & 4                             \\
Add Reference          & 184                            & 3                            & 1                                & 5                             \\
Find Non-Text Material & 163                            & 7                            & 2                                & 10                            \\
\textbf{Translate}     & \textbf{154}                   & \textbf{23}                  & \textbf{3}                       & \textbf{15}                   \\
Question Answering     & 52                             & 115                          & 1                                & 21                            \\
Send                   & 187                            & 10                           & 0                                & 2                             \\
Ideation               & 91                             & 36                           & 0                                & 14                            \\
Create Outline         & 144                            & 16                           & 0                                & 4                             \\
Collaboration          & 129                            & 20                           & 0                                & 5             \\
\bottomrule
\end{tblr}
\end{table*}

\begin{table*}[]
\caption{Results of coded ``Goals'' participants had for delegating different tasks to AI, with a focus on the capabilities summarize, extend, and translate. We discovered that participants aimed for efficiency, learning, and improving text when delegating a summarize task to AI. Delegating an extend task to AI they desire to gain inspiration. Translating occurs most often with the goal of efficiency.}
\Description{This table shows results of coded ``Goals'' participants had for delegating a certain AI capability. The table lists the AI capabilities and counts of coded goals, such as inspiration and efficiency.}
\label{tab:ai-skills-goals}
\small
\begin{tblr}{lrrrrr}
\toprule
\textbf{AI Capability}               & \SetCell[r=1]{l}{\textbf{Inspiration}} & \SetCell[r=1]{l}{\textbf{Efficiency}} & \SetCell[r=1]{l}{\textbf{Learning}} & \SetCell[r=1]{l}{\textbf{Improve Text}} & \SetCell[r=1]{l}{\textbf{Coordination}} \\ \hline
Revise                 & 9                               & 42                             & 21                           & 136                              & 8                                \\
\textbf{Summarize}     & \textbf{3}                      & \textbf{81}                    & \textbf{30}                  & \textbf{59}                      & \textbf{2}                       \\
Format                 & 0                               & 208                            & 1                            & 5                                & 0                                \\
Shorten                & 2                               & 114                            & 5                            & 52                               & 1                                \\
\textbf{Extend}        & \textbf{122}                    & \textbf{44}                    & \textbf{8}                   & \textbf{22}                      & \textbf{2}                       \\
Highlight              & 0                               & 176                            & 5                            & 27                               & 5                                \\
Cut                    & 0                               & 185                            & 1                            & 10                               & 0                                \\
Complete               & 156                             & 2                              & 2                            & 4                                & 0                                \\
Draft                  & 142                             & 131                            & 4                            & 2                                & 0                                \\
Inspire                & 144                             & 95                             & 12                           & 11                               & 2                                \\
Add Reference          & 0                               & 193                            & 0                            & 0                                & 0                                \\
Find Non-Text Material & 120                             & 101                            & 8                            & 3                                & 0                                \\
\textbf{Translate}     & \textbf{3}                      & \textbf{148}                   & \textbf{27}                  & \textbf{14}                      & \textbf{7}                       \\
Question Answering     & 3                               & 95                             & 87                           & 40                               & 10                               \\
Send                   & 0                               & 0                              & 0                            & 55                               & 199                              \\
Ideation               & 88                              & 6                              & 10                           & 41                               & 2                                \\
Create Outline         & 104                             & 113                            & 5                            & 3                                & 8                                \\
Collaboration          & 1                               & 4                              & 3                            & 10                               & 139   \\
\bottomrule
\end{tblr}
\end{table*}

\begin{table*}[]
\caption{Results of coded ``Roles'' in which participants saw the AI based on their message when delegating a task, with a focus on the capabilities summarize, extend, and translate (bold). We discovered that participants considered the AI most often as a co-author for all three tasks.}
\Description{This table shows results of coded ``Roles'' in which participants saw the AI based on their message when delegating a certain AI capability. The table lists the AI capabilities and counts of coded roles, such as co-author and advisor.}
\label{tab:ai-skills-roles}
\small
\begin{tblr}{lrrrr}
\toprule
\textbf{AI Capability}               & \SetCell[r=1]{l}{\textbf{Co-Author}} & \SetCell[r=1]{l}{\textbf{Advisor}} & \SetCell[r=1]{l}{\textbf{Assistant}} & \SetCell[r=1]{l}{\textbf{Editor}} \\ \hline
Revise                 & 122                           & 9                           & 44                            & 20                         \\
\textbf{Summarize}     & \textbf{121}                  & \textbf{2}                  & \textbf{33}                   & \textbf{4}                 \\
Format                 & 0                             & 0                           & 209                           & 0                          \\
Shorten                & 144                           & 0                           & 20                            & 9                          \\
\textbf{Extend}        & \textbf{149}                  & \textbf{3}                  & \textbf{13}                   & \textbf{4}                 \\
Highlight              & 10                            & 1                           & 170                           & 7                          \\
Cut                    & 4                             & 1                           & 187                           & 0                          \\
Complete               & 159                           & 0                           & 2                             & 2                          \\
Draft                  & 158                           & 4                           & 11                            & 1                          \\
Inspire                & 153                           & 5                           & 7                             & 14                         \\
Add Reference          & 0                             & 0                           & 192                           & 0                          \\
Find Non-Text Material & 19                            & 0                           & 161                           & 1                          \\
\textbf{Translate}     & \textbf{159}                  & \textbf{10}                 & \textbf{23}                   & \textbf{5}                 \\
Question Answering     & 35                            & 9                           & 128                           & 18                         \\
Send                   & 0                             & 1                           & 197                           & 0                          \\
Ideation               & 68                            & 24                          & 15                            & 38                         \\
Create Outline         & 141                           & 7                           & 12                            & 4                          \\
Collaboration          & 3                             & 2                           & 144                           & 6       \\
\bottomrule
\end{tblr}
\end{table*}

\end{document}